\documentclass[aps,nofootinbib,notitlepage,superscriptaddress]{revtex4-1}

\expandafter\let\csname equation*\endcsname\relax
\expandafter\let\csname endequation*\endcsname\relax
\expandafter\let\csname eqnarray*\endcsname\relax
\expandafter\let\csname endeqnarray*\endcsname\relax

\usepackage{amsmath}
\usepackage{amssymb}
\usepackage{graphicx}
\usepackage{braket}
\usepackage{color}
\usepackage{float}
\usepackage{soul}

\newcommand{\A}{\nonumber}
\newcommand{\RN}[1]{\textup{\uppercase\expandafter{\romannumeral#1}}}

\begin{document}
\setlength{\parindent}{0mm}

\title{The gravitational field of a laser beam beyond the short wavelength approximation}
\author{Fabienne Schneiter}
\email{fabienne.schneiter@uni-tuebingen.de }
\address{Eberhard-Karls-Universit\"at T\"ubingen, Institut f\"ur Theoretische Physik, 72076 T\"ubingen, Germany}

\author{Dennis R\"atzel}
\email{dennis.raetzel@univie.ac.at}
\affiliation{University of Vienna, Faculty of Physics, Boltzmanngasse 5, 1090 Vienna, Austria}

\author{Daniel Braun}
\address{Eberhard-Karls-Universit\"at T\"ubingen, Institut f\"ur Theoretische Physik, 72076 T\"ubingen, Germany}

\maketitle

\section*{Abstract}
\begin{center}

Light carries energy, and therefore, it is the source of a gravitational field. The gravitational field of a beam of light in the short wavelength approximation has been studied by several authors. In this article, we consider light of finite wavelengths by describing a laser beam as a solution of Maxwell's equations and taking diffraction into account. Then, novel features of the gravitational field of a laser beam become apparent, such as frame-dragging due to its spin angular momentum and the deflection of parallel co-propagating test beams that overlap with the source beam. Even though the effects are too small to be detected with current technology, they are of conceptual interest, revealing the gravitational properties of light.
 
\end{center}

\section{Introduction}

The gravitational field of a light beam has first been studied by Tolman, Ehrenfest and Podolski in 1931 \cite{tolman_1931}, who described the light beam as a one-dimensional "pencil of light". Later, a description for the gravitational field of a cylindrical beam of light of a finite radius has been presented by Bonnor \cite{bonnor_1969}. In this description, light has been modeled as a continuous fluid moving at the speed of light. A central feature of these two models is the lack of diffraction; the beams do not diverge. This corresponds to the short wavelength limit where all wavelike properties of light are neglected. Further studies of the gravitational field of light that share this feature include the investigation of two co-directed parallel cylindrical light beams of finite radius \cite{nackoney_1973,banerjee_1975}, spinning non-divergent light beams \cite{Vsevolodovich:1989the}, non-divergent light beams in the framework of gravito-electrodynamics \cite{Faraoni:1999grav}, and the gravitational field of a point like particle moving with the speed of light \cite{aichelburg_1971,Voronov1973}.

In contrast, the wavelike properties of light have been taken into account in \cite{vanholten_2011}, where the gravitational field of a plane electromagnetic wave has been investigated. An approach to take finite wavelengths into account for the case of a laser pulse has been given in \cite{raetzel_2016,Raetzel:2017grav}, where, however, diffraction has been neglected. In this article we describe the laser beam as a solution to Maxwell's equations. This is done perturbatively by an expansion in the beam divergence, which is considered to be small. The zeroth order of the expansion corresponds to the paraxial approximation and coincides with the result of \cite{bonnor_1969}. In the first order in the beam divergence, frame-dragging due to the spin angular momentum of circularly polarized beams occurs.  In the fourth order in the divergence angle, a parallel co-propagating test beam of light overlapping with the source laser beam is found to be deflected by the gravitational field of the laser beam.

The properties of light are inherent in modern physics. They were used to derive special and general relativity and they are often the basis for new approaches to spacetime theories. Furthermore, the gravitational field of laser beams is a phenomenon on the interface of general relativity and quantum mechanics as laser beams can be brought into non-classical states. For the progress of modern physics it is of great importance to study such phenomena, as they may give some insight into quantum gravity. Hence, it is necessary to study the gravitational properties of laser light in sufficient detail. In this article one of the most fundamental features of laser light, its wave properties, is taken into account for the first time. Therefore, even though the effects we present in this article are very small and not measurable with current technology, they are of general interest for the physics community.

We would like to point out that, if detection of the gravitational field of light may be feasable at some point in the future, it is very likely that strongly focussed laser beams will be involved in the corresponding experiments. However, due to the wavelike nature of light, there is a fixed relation between a laser beam's divergence angle and the width of its focus. This feature limits the experimental possibilities further. This has to be taken into account to obtain the sensitivity that would be necessary to detect the gravitational field of light at some point in the future. Therefore, future advanced detection schemes that may be promising to detect the gravitational field of light have to be assessed using the detailed description given in this article. Hence, this article is of importance to future considerations of the possibilities to detect the gravitational field of light.

We proceed as follows: In Sec. \ref{sec:beam}, we describe a focused laser beam as a solution to Maxwell's equations. This is done perturbatively, as an expansion in the small beam divergence angle $\theta$. Furthermore, we derive the energy-momentum tensor for a circularly polarized laser beam.
In Sec.~\ref{sec:lingrav}, we introduce the framework of linearized gravity. The equations determining the metric perturbation and solutions with Green's functions are given in Sec.~\ref{sec:metric}. Then we discuss the specific effects appearing in the different orders of the expansion in $\theta$ of the gravtational field: In Sec.~\ref{sec:leading}, we discuss the zeroth order, which corresponds to the paraxial approximation. Frame-dragging happens in the first order of the metric perturbation and is explained in Sec.~\ref{sec:firstorderframedrag}. That a co-propagating parallel light ray is deflected in the gravitational field of the laser beam is shown in Sec.~\ref{sec:fourthorderdefl}. Some conclusions are given in Sec.~\ref{sec:conclusions}.

Throughout the article we use the following notation: For spacetime coordinates  we use greek indices, like $x^\alpha$, and for spatial coordinates we use latin indices, like $x^k$. For the Minkowski metric, we choose the convention $\eta_{\alpha\beta}={\rm diag}(-1,1,1,1)$.

\section{Describing the Laser Beam}
\label{sec:beam}

In this section we describe the laser beam as a Gaussian beam, a perturbative solution to Maxwell's equations. The solution is expanded in the beam divergence, which is assumed to be small. Finding a solution for the vector potential, we calculate the energy-momentum tensor which will be used in the next section to determine the spacetime metric.

\subsection{The field strength tensor}
\label{sec:fieldstrength}

The laser beam is a monochromatic plane wave whose intensity distribution in the directions perpendicular to the direction of propagation decreases with a Gaussian factor. It is a perturbative solution of Maxwell's equations: an expansion in the beam divergence, the opening angle of the beam, which is assumed to be small. This solution is obtained by making the ansatz that the vector potential is a plane wave enveloped by a function depending on the position.

More specifically, the vector potential of the Gaussian beam is obtained as follows: It has to satisfy Maxwell's equations in form of the wave equations, 
\begin{eqnarray}\label{m}
	\Box  A_\alpha(t,x,y,z)
	&=& 0\;, 
\end{eqnarray}		
where $\Box = \eta^{\alpha\beta}\partial_\alpha\partial_\beta= -\frac{1}{c^2}\partial_t^2+\partial_x^2+\partial_y^2+\partial_z^2$ is the d'Alembert operator and we choose the Lorenz gauge condition $\eta^{\alpha\beta}\partial_\alpha A_\beta=0$. For convenience, we work in the dimensionless coordinates $\tau=\frac{ct}{w_0}$, $\xi=\frac{x}{w_0}$, $\chi=\frac{y}{w_0}$, $\zeta=\frac{z}{w_0}$, where $w_0$  is the beam waist. Writing $\{  x^\alpha \}$ for the coordinates $\{ct,x,y,z\}$ and $\{x^{\bar\alpha}\}$ for the coordinates $\{\tau,\xi,\chi,\zeta\}$, we obtain for the Minkowski metric
\begin{equation}\label{min}
	\eta_{\bar\alpha\bar\beta}
	= \frac{dx^\alpha}{dx^{\bar\alpha}}\;\frac{dx^\beta}{dx^{\bar\beta}}\;\eta_{\alpha\beta}
	=	w_0^2\;{\rm diag}\left(-1,1,1,1\right)\;.
\end{equation}
The vector potential transforms as $A_{\bar\alpha}=\frac{dx^\alpha}{dx^{\bar\alpha}}A_\alpha$.
We make the ansatz that the vector potential is monochromatic and can be written as
\begin{eqnarray}\label{an}
	A_{\bar\alpha}(\tau,\xi,\chi,\zeta)=\mathcal{A} v_{\bar\alpha}(\xi,\chi,\theta\zeta)e^{i\frac{2}{\theta}(\zeta-\tau)}\;,
\end{eqnarray}	
where $\theta=2/(w_0k)$ is the divergence angle of the beam, $k$ is the wave vector and $\mathcal{A}$ is the amplitude. The vector envelope function $v_{\bar\alpha}$ is assumed to depend on $\zeta$ only through the combination $\theta\zeta$. With the ansatz (\ref{an}), we obtain the Helmholtz equation for the envelope function
\begin{eqnarray}\label{holtz}
	\left(\partial_\xi^2+\partial_\chi^2+\theta^2 \partial_{\theta\zeta}^2+ 4i \partial_{\theta\zeta}\right)v_{\bar\alpha}(\xi,\chi,\theta\zeta)
	&=& 0\;.
\end{eqnarray}
We consider $\theta$ to be small, which implies that the envelope function changes much more slowly in $z$-direction than in $x$-direction or in $y$-direction. Then, we make the ansatz that $v_{\bar\alpha}$ can be written as a power series of $\theta$, \footnote{An expansion in orders of $\theta^2$ has been presented by Davis \cite{Davis:1979theo}. Here, we consider the general expansion to allow for helicity eigenstates later on.}
\begin{eqnarray}\label{eq:expansionv}
	v_{\bar\alpha}(\xi,\chi,\theta\zeta)
	&=& \sum_{n=0}^\infty \theta^{n}v^{(n)}_{\bar\alpha}(\xi,\chi,\theta\zeta)\;,
\end{eqnarray}
where $v_{\bar\alpha}^{(n)}$ are the coefficients in the power series. The Helmholtz equation (\ref{holtz}) leads to the differential equations
\begin{eqnarray}\label{eq:parhelm}
	\left(\partial_\xi^2+\partial_\chi^2+ 4i\partial_{\theta\zeta}\right)v^{(0)}_{\bar\alpha}(\xi,\chi,\theta\zeta)
	&=& 0\;,\\
	\label{eq:parhelm1}
	\left(\partial_\xi^2+\partial_\chi^2+ 4i\partial_{\theta\zeta}\right)v^{(1)}_{\bar\alpha}(\xi,\chi,\theta\zeta)
	&=& 0\;,\\	\label{eq:parhelmhigher}\left(\partial_\xi^2+\partial_\chi^2+ 4i\partial_{\theta\zeta}\right)v^{(n)}_{\bar\alpha}(\xi,\chi,\theta\zeta)
	&=&	-\partial_{\theta\zeta}^2v^{(n-2)}_{\bar\alpha}(\xi,\chi,\theta\zeta)\,,\;\rm{for}\,\,n> 1\,.
\end{eqnarray}
Note, that this set of equations couples components of $v_{\bar\alpha}$ of odd $n$ to other components of odd $n$ and components with even $n$ to other components of even $n$. Therefore, we obtain two independent hierarchies of components of $v_{\bar\alpha}$. We will couple odd and even components later when we introduce helicity. 

Eq.~(\ref{eq:parhelm}) is known as the paraxial Helmholtz equation. It can be interpreted as a Schr\"odinger equation in two spatial dimensions with $m/\hbar = 2$ when $\theta\zeta$ is seen as a time variable, i.e.
\begin{eqnarray}\label{eq:parhelmschroe}
	i\partial_{\theta\zeta}v^{(0)}_{\bar\alpha}(\xi,\chi,\theta\zeta)
	&=& -\frac{1}{4}\Delta_{2d} v^{(0)}_{\bar\alpha}(\xi,\chi,\theta\zeta)\;,
\end{eqnarray}
where $\Delta_{2d}=\partial_\xi^2+\partial_\chi^2$ is the two dimensional Laplace operator. A solution of Eq.~(\ref{eq:parhelmschroe}) has to spread similar to the wave packet of a massive particle in quantum mechanics. Here, the spreading of the wave packet corresponds to the divergence of the beam. The solution of Eq.~(\ref{eq:parhelmschroe}) that we are interested in is a Gaussian wave packet. Furthermore, we want the wave packet to be centered on the optical axis and that it is rotationally symmetric about the optical axis. With these conditions, we obtain for the lowest order
\begin{equation}\label{eq:zerothorder}
	v^{(0)}_{\bar\alpha}(\xi,\chi,\theta\zeta)
	= \epsilon^{(0)}_{\bar\alpha}v_0(\xi,\chi,\theta\zeta)\;,
\end{equation}	
where the function $v_0$ is given by
\begin{equation}	
	v_0(\xi,\chi,\theta\zeta) = \mu(\theta\zeta)e^{-\mu(\theta\zeta)\rho^2}\;,\;
\end{equation}
and where $\rho=\sqrt{\xi^2+\chi^2}$, $\epsilon^{(0)}_{\bar\alpha}$ is the constant polarization co-vector and $\mu(\theta\zeta)=1/(1+i\theta\zeta)$ relates the spread of the Gaussian wave packet and the divergence angle of the beam. Eq.~(\ref{eq:zerothorder}) represents the Gaussian beam in lowest order in the divergence angle $\theta$. A graphic representation can be found in Fig.~1.
\begin{figure}[H]\center\label{fig:paysages}
	\includegraphics[scale=0.7]{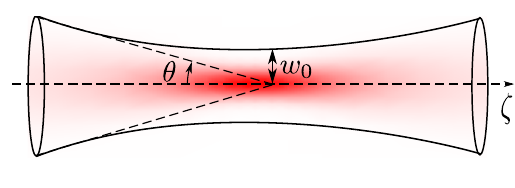}
	\caption{Schematic illustration of the Gaussian beam, the beam waist $w_0$, the Rayleigh length $z_R$ and the beam divergence $\theta$. More specifically, the figure illustrates the scalar envelope function $v_0$ of the vector potential of the Gaussian beam in a plane that contains the optical axis (represented by the dashed horizontal line). Due to the rotational symmetry of the envelope function about the optical axis, the vertical axis can be any direction transversal to the optical axis. The thick curved lines mark the distance $w(\zeta)=1/|\mu(\theta\zeta)|$ from the optical axis at which the absolute value of the envelope function reaches $1/e$ times its maximum.}
\end{figure}
The first order solution fulfills the same paraxial Helmholtz equation as the zeroth order solution. Therefore, we set
\begin{eqnarray}\label{eq:firstorder}
	v^{(1)}_{\bar\alpha}(\xi,\chi,\theta\zeta)
	&=& \epsilon^{(1)}_{\bar\alpha}v_0(\xi,\chi,\theta\zeta)\;.
\end{eqnarray}
The equations for the higher order terms in Eq.~(\ref{eq:parhelmhigher}) correspond to Schr\"odinger equations with an additional term proportional to the solution of the equation two orders lower, which has the effect of a source term,
\begin{eqnarray}
	\label{eq:parhelmhigherschroe} i\partial_{\theta\zeta} v^{(n)}_{\bar\alpha}(\xi,\chi,\theta\zeta)
	&=&	- \frac{1}{4}\Delta_{2d} v^{(n)}_{\bar\alpha}(\xi,\chi,\theta\zeta) - \frac{1}{4}\partial_{\theta\zeta}^2 v^{(n-2)}_{\bar\alpha}(\xi,\chi,\theta\zeta)\,,\;\;\;\rm{for}\,\,n\geq 1\,.
\end{eqnarray}
Finally, we have to specify the polarization co-vectors $\epsilon_{\bar \alpha}$ and the terms in the expansion of the envelope function of even $n$. We will do so for a Gaussian beam of circular polarization in the following. First, note that the components of the vector potential are not independent; the Lorenz gauge condition we imposed leads to
\begin{equation}\label{eq:lorenzAtau}
	 A_{\tau} = \frac{i\theta}{2} \partial_{\tau} A_{\tau} =  \frac{i\theta}{2} \left(\partial_\xi A_\xi + \partial_\chi A_\chi + \theta\partial_{\theta\zeta} A_{\zeta} \right)  \,.
\end{equation}
With this identity, $A_{\tau}$ can be eliminated from the space-time components of the field strength tensor $F_{\bar\alpha\bar\beta}=\partial_{\bar\alpha} A_{\bar\beta} - \partial_{\bar\beta} A_{\bar\alpha}$ as
\begin{eqnarray}
	F_{\tau\bar{a}} &=& -F_{\bar{a}\tau} = -\frac{2i}{\theta} A_{\bar{a}} - \frac{i\theta}{2} 
	\delta^{\bar b \bar c}\partial_{\bar a}\partial_{\bar{b}} A_{\bar{c}}\;,
\end{eqnarray}
where $\delta^{\bar b\bar c}$ is the Kronecker delta. As the vector potential, the field strength tensor can be expanded as 
\begin{equation}\label{eq:expf}
	F_{\bar\alpha\bar\beta} 	= \sum_{n=0}^\infty \theta^{n} \frac{w_0E_0}{\sqrt{2}} f_{\bar\alpha\bar\beta}(\xi,\chi,\theta\zeta)e^{i\frac{2}{\theta}(\zeta-\tau)}\,,
\end{equation}
where $E_0=\sqrt{2}\mathcal{A}/(w_0\theta)$ and a direct relation between $v_{\bar\alpha}^{(n)}$ and $f^{(n)}_{\bar\alpha\bar\beta}$ can be established, which is given in Appendix \ref{sec:fexpansion}.

\subsubsection{Circularly polarized beams}
\label{sec:circpolbeam}

In the last step, we have to specify the polarization of the beam that we want to consider. In this article, we will focus on circularly polarized beams. We define a circularly polarized beam as a helicity state which is an eigenstate of the generator of the duality transformations $F'_{\bar{\alpha}\bar{\beta}} =  F_{\bar{\alpha}\bar{\beta}}\cos{\varphi} + \star F_{\bar{\alpha}\bar{\beta}}\sin{\varphi}$, where $\star F_{\bar{\alpha}\bar{\beta}} = \frac{1}{2}\sqrt{-\rm{det}(\eta)}\epsilon_{\bar{\alpha}\bar{\beta}\bar{\gamma}\bar{\delta}}F^{\bar{\gamma}\bar{\delta}}$ is the Hodge dual of $F_{\bar{\alpha}\bar{\beta}}$ and $\epsilon_{\bar{\alpha}\bar{\beta}\bar{\gamma}\bar{\delta}}$ is the completely anti-symmetric Levi-Civita symbol with $\epsilon_{0123}=-1$. The invariance of Maxwell's equations under these duality transformations and the corresponding conservation laws were worked out in \cite{Calkin:1965ani}. The generator of the duality transformation $D_\theta = \exp(i\varphi\Lambda): F_{\bar{\alpha}\bar{\beta}} \mapsto F'_{\bar{\alpha}\bar{\beta}}$ is $\Lambda: F_{\bar{\alpha}\bar{\beta}} \mapsto -i\,\star F_{\bar{\alpha}\bar{\beta}}$ since $\star\star F_{\bar{\alpha}\bar{\beta}} = - F_{\bar{\alpha}\bar{\beta}}$. 

The field tensors corresponding to the vector potentials of well-defined helicity are eigenstates of $\Lambda$ with eigenvalues $\lambda=\pm 1$. There are two options to obtain these eigenstates. One option is to start with a helicity eigenstate of zeroth order in $\theta$, construct the corresponding higher order terms of the expansion of the envelope function of even $n$ with Eq.~(\ref{eq:parhelmhigherschroe}), obtain the odd terms in the expansion of the envelope function with the Lorenz gauge condition in Eq.~(\ref{eq:lorenzAtau}), calculate the field strength tensor and project it with $(1+\lambda\Lambda)/2$. This option is presented in Appendix \ref{sec:projsol}. 

In the main text of this article, we follow the second option, where a vector potential is constructed order by order by taking into account the condition $(1-\lambda\Lambda)F^\lambda_{\bar{\alpha}\bar{\beta}}=0$ and the expansion in Eq.~(\ref{eq:expf}) in each order separately. This construction is presented in Appendix \ref{sec:fexpansion}. Starting from $v^{(0)}_{\bar\alpha} = \epsilon_{\bar\alpha}^{(0)} v_0$, where $\epsilon_{\bar a}^{(0)}=w_0(1,-\lambda i,0)/\sqrt{2}$ and $\bar a \in \{\xi,\chi,\zeta\}$, and taking the solutions of even orders from \cite{Salamin:2007fie} into account, we obtain
\begin{eqnarray}\label{eq:circpolenv}
	v^{\lambda (0)}_{\bar a}
	&=& \epsilon^{(0)}_{\bar a}v_0\;,\\
	 v^{\lambda (1)}_{\bar a}
	&=& -\epsilon^{(1)}_{\bar a} \frac{i\mu}{2\sqrt{2}}\left(\xi - i\lambda\chi\right)v_0\;,\\
	v^{\lambda (2)}_{\bar a}
	&=&	\frac{\mu}{2}\left(1 - \frac{1}{2}\mu^2\rho^4 \right)v^{\lambda (0)}_{\bar a}\;,\\
	v^{\lambda (3)}_{\bar a}
	&=&	\frac{\mu}{4}\left(4 +  \mu\rho^2 - \mu^2\rho^4 \right)v^{\lambda (1)}_{\bar a}\;,\\
	v^{\lambda (4)}_{\bar a}
	&=&	\frac{\mu^2}{16}  \left(6  - 3\mu^2\rho^4 - 2\mu^3\rho^6 + \frac{1}{2}\mu^4 \rho^8 \right)v^{\lambda (0)}_{\bar a} \;,
\end{eqnarray}
where $\epsilon^{(1)}_{\bar a}=w_0(0,0,1)$. The corresponding vector potential is given as 
$A^\lambda_{\bar\alpha}	= \sum_{n=0}^4 \theta^{n} \mathcal{A} v^{\lambda(n)}_{\bar\alpha}(\xi,\chi,\theta\zeta) e^{i\frac{2}{\theta}(\zeta-\tau)}$, where the component $A^\lambda_{\tau}$ is given through the Lorenz gauge condition in Eq.~(\ref{eq:lorenzAtau}). Linearly polarized Gaussian beams are obtained as linear combinations of helicity eigenstates; for example, $A^\xi_{\bar \alpha}:=(A^+_{\bar \alpha}+A^-_{\bar \alpha})/\sqrt{2}$ is the vector potential of a laser beam that is linearly polarized in the $\xi$-direction. Note that all contributions to $v_\alpha$ except the leading order decay faster than $v^{\lambda (0)}_{\bar a}$ for $\theta\zeta\rightarrow \infty$. Hence, $v_{\bar a}\approx v^{\lambda (0)}_{\bar a}$ for large $\theta\zeta$. 

\subsection{Three distinct scenarios}

The beam divergence $\theta$, which is assumed to be small, is related to the wave vector $k$, the beam waist $w_0$ and the Rayleigh length $z_R$ through
\begin{eqnarray}
	k
	= \frac{2}{w_0\theta}
	= \frac{2}{z_R \theta^2}\;.
\end{eqnarray}
The beam waist $w_0$ describes the width of the beam at its focal point, i.e.~at $\zeta=0$, and the Rayleigh length is the distance from the focal point along the direction of propagation such that the cross section of the beam is doubled, as illustrated in Fig. 1.
There are basically three scenarios for which the condition that $\theta$ is small is satisfied:
\begin{enumerate}
	\item $k={\rm constant}$: If the wave vector $k$ is kept constant, 		the beam waist $w_0$ and the Rayleigh length $z_R$ have to be large, and $z_R\gg w_0$ has to hold. Keeping the wave vector constant is the characteristic feature of a plane wave. 
	If the beam is very long, it may be compared to the infinitely extended plane waves, which are described by the pp-wave metrics.\footnote{ See chapter 35 in \cite{mtw}.}  
	\item $w_0={\rm constant}$: Keeping the beam waist $w_0$ fixed, the wave vector $k$ and the Rayleigh length $z_R$ have to be large, and in addition we find $z_R\gg \frac{1}{k}$. This situation describes an almost parallel beam of a given waist. If the beam is very long and the beam waist is considered to be small, such that it is approximately a cylinder of light, it may be compared to the solution found by Bonnor \cite{bonnor_1969} for an infinitely long cylinder of light.
	\item $z_R={\rm constant}$: Keeping the Rayleigh length fixed, the wave vector $k$ has to be large and the beam waist $w_0$ has to be small. This case corresponds to a very thin and almost parallel beam along the $z$-axis, whose energy-density is accordingly high. This is the solution given by Tolman, Ehrenfest and Podolski \cite{tolman_1931}.
\end{enumerate}
In the following, we will keep the beam waist $w_0$ constant. 

\subsection{The energy-momentum tensor}

To derive the gravitational field of the laser beam, we have to derive its energy-momentum tensor first. Let us define the real part of $F_{\bar{\alpha}\bar{\beta}}$ as $\mathrm{Re}(F)_{\bar{\alpha}\bar{\beta}}$. In terms of $\mathrm{Re}(F)_{\bar{\alpha}\bar{\beta}}$, the energy-momentum tensor is defined as 
$	T_{\bar\alpha\bar\beta}
	=	c^2\varepsilon_0({\mathrm{Re}(F)_{\bar\alpha}}^{\bar\sigma}\mathrm{Re}(F)_{\bar\beta\bar\sigma}
		-\frac{1}{4}\eta_{\bar\alpha\bar\beta}\mathrm{Re}(F)^{\bar\delta\bar\rho}\mathrm{Re}(F)_{\bar\delta\bar\rho})$.
Therefore, the energy-momentum tensor can be decomposed into the real term 
\begin{eqnarray} 
	{(T^r)}_{\bar{\alpha}\bar{\beta}} = \frac{c^2\varepsilon_0}{2}\mathrm{Re}\left({F_{\bar\alpha}}^{\bar\sigma}F^*_{\bar\beta \bar\sigma} -\frac{1}{4}\eta_{\bar\alpha\bar\beta}F^{\bar\delta\bar\rho}F^*_{\bar \delta\bar\rho} \right)\;,
\end{eqnarray}
the complex term
\begin{eqnarray}
{(T^c)}_{\bar{\alpha}\bar{\beta}} = \frac{c^2\varepsilon_0}{4}\left({F_{\bar\alpha}}^{\bar\sigma}F_{\bar\beta\bar\sigma}		-\frac{1}{4}\eta_{\bar\alpha\bar\beta}F^{\bar\delta\bar\rho}F_{\bar\delta\bar \rho}\right)\;,
\end{eqnarray}
and its complex conjugate $(T^c)_{\bar{\alpha}\bar{\beta}}^*$. The term $(T^c)_{\bar{\alpha}\bar{\beta}}$ is highly oscillating with $i(\zeta-\tau)/\theta$ while these oscillations cancel in $(T^r)_{\bar{\alpha}\bar{\beta}}$. For eigenstates of the helicity operator with eigenvalue $\lambda=\pm 1$, the highly oscillating terms in $(T^c)_{\bar{\alpha}\bar{\beta}}$ and its complex conjugate vanish and it remains $T_{\bar{\alpha}\bar{\beta}} = (T^r)_{\bar{\alpha}\bar{\beta}}$. Therefore, the highly oscillating parts of the energy-momentum tensor can be interpreted as a result of the interference of contributions of different helicity in the field strength that come into play for linear or elliptical polarization. In the following, we will only consider circular polarization.

The components of the energy-momentum tensor are directly related to the energy density $\mathcal{E}^\lambda$, the Poynting vector $\vec S^\lambda$ and the Maxwell stress tensor $\sigma^\lambda_{ij}$ of the electromagnetic field,
\begin{eqnarray}\label{eq:Tmunured}
	T^\lambda_{\bar\alpha\bar\beta} &=&\begin{pmatrix}
	\mathcal{E}^\lambda & -S^\lambda_\xi/c & -S^\lambda_\chi/c & -S^\lambda_\zeta/c\\
	-S^\lambda_\xi/c & \sigma^\lambda_{11}&\sigma^\lambda_{12} &\sigma^\lambda_{13}\\
	-S^\lambda_\chi/c & \sigma^\lambda_{12} & \sigma^\lambda_{22} & \sigma^\lambda_{23}\\
	-S^\lambda_\zeta/c & \sigma^\lambda_{13} & \sigma^\lambda_{23} & \sigma^\lambda_{33}
	\end{pmatrix}\;.
\end{eqnarray}
For the field strength tensor $F^{\lambda}_{\bar{\alpha}\bar{\beta}}=\partial_{\bar\alpha}A^\lambda_{\bar\beta}-\partial_{\bar\beta}A^\lambda_{\bar\alpha}$ of a circularly polarized laser  beam which we specified in Sec.~\ref{sec:circpolbeam}, the energy density, the Poynting vector and the stress tensor components are given in Appendix \ref{sec:pointing}.

The power transmitted in the direction of propagation is given by  $P=\int_0^{2\pi}d\phi\int_0^\infty d\rho \,\rho S_\zeta$. In the leading order in the expansion in $\theta$, we obtain $P_0=\pi c\varepsilon_0 E_0^2 w_0^2/2$, where $E_0$ is the amplitude of the electric field in the leading order at the beamline. We may then express the amplitude in terms of the power as $E_0=\sqrt{\frac{2P_0}{\pi c\varepsilon_0 w_0^2}}$. For a power of $P_0\sim 10^{15}{\rm W}$ and a beam waist of $w_0\sim 10^{-3}{\rm m}$, the amplitude is $E_0\sim 10^{12}\;{\rm \frac{V}{m}}$.

As the field strength tensor, the energy-momentum tensor can be expanded in orders of $\theta$ as $T^\lambda_{\bar{\alpha}\bar{\beta}}=\sum_n \theta^n t_{\bar{\alpha}\bar{\beta}}^{\lambda(n)}$. Then, the gravitational field of the laser  beam can be calculated for each order and effects of different orders can be identified. We will present this analysis up to fourth order in $\theta$ in the following sections.

\section{Linearized gravity}
\label{sec:lingrav}

Assuming that the energy of the laser  beam is sufficiently small, we use the linearized theory of general relativity\footnote{ See chapter 18 in\cite{mtw}.} to describe its gravitational field. In Appendix \ref{sec:validity}, we make a rough estimation to show that this is reasonable. The metric $g_{\alpha\beta}$ consists of the metric for flat spacetime $\eta_{\alpha\beta}$ plus a small perturbation $h_{\alpha\beta}$ with $|h_{\alpha\beta}|\ll1$, 
\begin{eqnarray}\label{krokussli}
	g_{\alpha\beta}
	&=& \eta_{\alpha\beta}+h_{\alpha\beta}\;.
\end{eqnarray}
Therefore one neglects terms quadratic in the metric perturbation. In this case, one sees that the inverse of the metric reads $g^{\alpha\beta}=\eta^{\alpha\beta}-h^{\alpha\beta}$. The Einstein equations can be simplified to a set of linear equations in the metric perturbation. As the full general relativity has an invariance under coordinate transformation, its linearized approximation is invariant under linear coordinate transformations $x^\alpha\rightarrow {\tilde x}^\alpha = x^\alpha+\xi^\alpha$, where the metric perturbation transforms as $h_{\alpha\beta}\rightarrow \tilde h_{\alpha\beta}=h_{\alpha\beta}-\partial_\alpha \xi_\beta-\partial_\beta\xi_\alpha$.\footnote{It is assumed that $|\partial_\alpha \xi_\beta|$ is of the same order of magnitude as $h_{\alpha\beta}$.} Since curvature is described by the second derivatives of the metric, quantities depending on the curvature are invariant under linear coordinate transformations. 

To derive the linearized version of the Einstein equations, we assume the Lorenz gauge condition, 
$
	\partial^\alpha h_{\alpha\beta}
	= \partial_\beta {h_\alpha}^\alpha/2
$. The energy-momentum tensor has to be conserved, $\eta^{\alpha\beta}\partial_\alpha T_{\beta\gamma}=0$, which implies that the continuity equation is satisfied \cite{raetzel_2016,maggiore}.
The remaining gauge freedom is given by linear coordinate transformations $\xi_\alpha$ that satisfy $\Box\xi_\alpha=0$. Taking into account that the trace of the energy-momentum tensor ${T_\sigma}^\sigma$ is identically zero for the electromagnetic field, we obtain the linearized Einstein equations\footnote{ See Eq.~18.8b in \cite{mtw}.}
\begin{eqnarray}\label{eq:wave}
	\Box  h_{\alpha\beta}
	&=&	-\kappa T_{\alpha\beta}\;,
\end{eqnarray}
where $\kappa=16\pi G/c^4$ and $G$ is Newton's constant.  

In general relativity, coordinates have no physical meaning. Since the values of the components of the metric tensor depend on the choice of coordinates, we cannot extract physical information directly from them. Therefore, we have to investigate effects on test particles to learn about the gravitational field. The motion of test particles is governed by the geodesic equation 
\begin{eqnarray}\label{apfel}
	\frac{d^2\gamma^\mu}{d\varrho^2}
	&=&	-\Gamma^\mu_{\nu\rho} 
		\frac{d\gamma^\nu}{d\varrho}\frac{d\gamma^\rho}{d\varrho}\;,
\end{eqnarray}
where, in linearized gravity, the Christoffel symbols are given as
\begin{equation}\label{eq:gammadef}
	\Gamma^\mu_{\nu\rho} = \frac{1}{2}\eta^{\mu\sigma}\left(\partial_\nu h_{\sigma\rho} + \partial_\rho h_{\sigma\nu} - \partial_\sigma h_{\nu\rho}\right)\;.
\end{equation} 

\begin{figure}[H]\center\label{fig:champs}
	\includegraphics[scale=0.25]{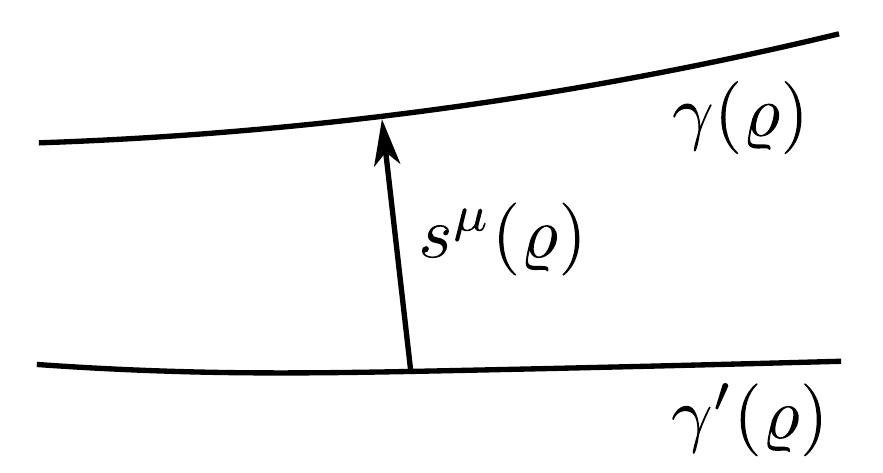}
	\caption{{Schematic illustration of the geodesic deviation equation: Two nearby geodesics $\gamma(\varrho)$ and $\gamma'(\varrho)$ are seperated by the vector $s^\mu(\varrho)$.} }
\end{figure}

A more direct way to analyse gravitational effects is through the spread and the contraction of the trajectories of test particles. This way, the test particles serve as each others reference. 
The relative acceleration between {two infinitesimally close geodesics $\gamma(\varrho)$ and $\gamma'(\varrho)$ parameterized by $\varrho$ is given by the geodesic deviation equation
\begin{equation}\label{eq:geodesicdev}
	a^\mu=\frac{D^2s^\mu}{d\varrho^2}={R^{\mu}}_{\rho\sigma\alpha}(\gamma)\dot{\gamma}^\rho \dot{\gamma}^\sigma s^\alpha\,,
\end{equation}
where $s$ is the separation vector between the geodesics,} $D/d\varrho=\dot\gamma^\mu \nabla_\mu$ is the covariant derivative along the geodesic $\gamma(\varrho)$ and ${R^{\mu}}_{\rho\sigma\alpha}$ is the Riemann curvature tensor. This is illustrated in Fig.~\ref{fig:champs}.
 In the linearized theory, the pulled down Riemann curvature tensor is given by
\begin{equation}\label{eq:R}
	R_{\alpha\beta\gamma\delta}
	=	\frac{1}{2}\left(\partial_\beta\partial_\gamma h_{\delta\alpha}
		-\partial_\beta\partial_\delta h_{\gamma\alpha}-\partial_\gamma\partial_\alpha h_{\beta\delta}
		+\partial_\delta\partial_\alpha h_{\beta\gamma}\right)\;	.	
\end{equation}
Since the metric perturbation transforms as $h_{\alpha\beta}\rightarrow \tilde h_{\alpha\beta}=h_{\alpha\beta}-\partial_\alpha \xi_\beta-\partial_\beta\xi_\alpha$, we find that $R_{\alpha\beta\gamma\delta}$ is invariant under a linearized coordinate transformation. 

\section{The Metric of the laser beam}
\label{sec:metric}

Solving Eq.~(\ref{eq:wave}) for the energy-momentum tensor (\ref{eq:Tmunured}) with emitter and absorber\footnote{ In this article, emitter and absorber always refers to the emitter and the absorber of the source laser beam for the case of a finitely extended beam. The emitter can be associated with the laser resonator and the active material and the absorber can be imagined as a beam dump.} at general positions can be quite cumbersome. In the following, we will consider two different limiting situations instead; we consider the case of the distance between emitter and absorber of the laser beam being very large and very small.  

In the first situation, we can neglect the rapid change of the field strength at the emitter and the absorber of the laser beam. Then we can take into account that $T_{\bar\alpha\bar\beta}$ is changing slowly in $\zeta$. In particular, we have $T^\lambda_{\bar\alpha\bar\beta}=\bar{T}^\lambda_{\bar\alpha\bar\beta}(\xi,\chi,\theta\zeta)$. Therefore, we can expand the metric perturbation similar to Eq.~(\ref{eq:expansionv}) as
\begin{eqnarray}\label{eq:expansionh}
	h^\lambda_{\bar\alpha\bar\beta}(\xi,\chi,\theta\zeta)
	&=& \sum_{n=0}^\infty \theta^{n}h_{\bar\alpha\bar\beta}^{\lambda(n)}(\xi,\chi,\theta\zeta)\;,
\end{eqnarray}
and the linearized Einstein equations (\ref{eq:wave}) lead to the differential equations
\begin{eqnarray}\label{eq:poissonh0}
	\Delta_{2d} h_{\bar\alpha\bar\beta}^{\lambda(0)}
	&=& -\kappa w_0^2\, \bar t^{\lambda(0)}_{\alpha\beta}\;,\\
	\label{eq:poissonh1}
	\Delta_{2d} h_{\bar\alpha\bar\beta}^{\lambda(1)}
	&=& -\kappa w_0^2 \, \bar t^{\lambda(1)}_{\bar\alpha\bar\beta}\;,\\
	\label{eq:poissonhhigher}\Delta_{2d} h_{\bar\alpha\bar\beta}^{\lambda(n)}
	&=&	-\kappa w_0^2\, \bar t^{\lambda(n)}_{\bar\alpha\bar\beta} - \partial_{\theta\zeta}^2 h_{\bar\alpha\bar\beta}^{\lambda(n-2)}\,,\;\rm{for}\,\,n > 1\,.
\end{eqnarray}
The solutions $h^{\lambda(n)}_{\bar\alpha\bar\beta}$ of Eqs.~(\ref{eq:poissonh0}), (\ref{eq:poissonh1}) and (\ref{eq:poissonhhigher}) can be given by using the free space Green's function for the Poisson equation in two dimensions as
\begin{equation}\label{eq:solpoisson}
h^{\lambda(n)}_{\bar\alpha\bar\beta}(\xi,\chi,\theta\zeta)=\frac{1}{4\pi}\int_{-\infty}^\infty d\xi'd\chi' 
\,{\rm log}\left((\xi-\xi')^2+(\chi-\chi')^2\right)
Q^{\lambda(n)}_{\bar\alpha\bar\beta}(\xi',\chi',\theta\zeta)\,,
\end{equation} 
where the $Q^{\lambda(n)}$ are the right hand sides of
Eqs.~(33), (34) and
(35) for $n=0,1$ or $n>1$, respectively,
and the $\bar{t}^{\lambda(n)}$ are defined through the expansion 
\begin{eqnarray}\label{eq:expansionh}
	\bar T^\lambda_{\bar\alpha\bar\beta}(\xi,\chi,\theta\zeta)
	&=& \sum_{n=0}^\infty \theta^{n}\bar
            t_{\bar\alpha\bar\beta}^{\lambda(n)}(\xi,\chi,\theta\zeta)\,. 
\end{eqnarray} The form of the solutions in Eq.~(\ref{eq:solpoisson}) was fixed by an additional condition that we did not discuss yet; we want the components of the Riemann curvature tensor to vanish at infinite distance from the beamline. As stated in Sec.~\ref{sec:lingrav}, the Riemann curvature tensor governs the spread and the contraction of the trajectories of test particles. This means, if the Riemann tensor vanishes, parallel geodesics stay parallel and there is no physical effect as the only reference for a test particle in linearized gravity can be another test particle. We can assume that there is no gravitational effect for infinite spatial distances from the beamline. Therefore, we assume that the Riemann curvature tensor ${R^{\mu}}_{\rho\sigma\alpha}$ vanishes for $\rho\rightarrow \infty$. The full discussion of the curvature condition and its implications are given in Appendix \ref{sec:mpert}. Additionally, Appendix \ref{sec:mpert} contains expressions for the components of the metric perturbation up to third order in $\theta$.

As we did before for the vector potential, the field strength tensor, the energy-momentum tensor and the metric perturbation, we expand the Christoffel symbols and the Riemann tensor in orders of $\theta$,
\begin{eqnarray}\label{eq:expansionGamma}
	(\Gamma^{\lambda})^{\bar\alpha}_{\bar\beta\bar\gamma}(\xi,\chi,\theta\zeta)
	&=& \sum_{n=0}^\infty \theta^{n} (\gamma^{\lambda(n)})^{\bar\alpha}_{\bar\beta\bar\gamma}(\xi,\chi,\theta\zeta)\;,
\end{eqnarray}
and
\begin{eqnarray}\label{eq:expansionR}
	R^\lambda_{\bar\alpha\bar\beta\bar\gamma\bar\delta}(\xi,\chi,\theta\zeta)
	&=& \sum_{n=0}^\infty \theta^{n} r^{\lambda(n)}_{\bar\alpha\bar\beta\bar\gamma\bar\delta}(\xi,\chi,\theta\zeta)\;,
\end{eqnarray}
respectively. With Eqs.~(\ref{eq:R}), (\ref{eq:gammadef}) and (\ref{eq:expansionh}), we can derive direct relations between the terms of the expansions $r^{\lambda(n)}_{\bar\alpha\bar\beta\bar\gamma\bar\delta}$ and $(\gamma^{\lambda(n)})^{\bar\alpha}_{\bar\beta\bar\gamma}$ and terms in the expansion of the metric perturbation $h_{\alpha\beta}^{\lambda(n)}$. They are given in Appendix \ref{sec:Rexpansion}.

\subsubsection{Small distance between emitter and absorber}

In the second situation, where we assume a short distance between emitter and absorber of the laser beam, the rapid change of the field strength at emitter and absorber of the laser beam cannot be neglected. Then, we solve the Einstein equations (\ref{eq:wave}) by use of their retarded solution
\begin{eqnarray}\label{eq:primeli}
	 h^\lambda_{\bar\alpha\bar\beta}(\tau,\xi,\chi,\zeta)
	&=&	\frac{\kappa w_0^2}{4\pi}\int_{-\infty}^\infty d\xi' d\chi' d\zeta' 
		\frac{T^\lambda_{\bar\alpha\bar\beta}\left(\tau -\sqrt{(\xi-\xi')^2+(\chi-\chi')^2+(\zeta-\zeta')^2},\xi',\chi',\theta\zeta'\right)}{\sqrt{(\xi-\xi')^2+(\chi-\chi')^2+(\zeta-\zeta')^2}}\;,
\end{eqnarray}
Furthermore, we can set $\theta\zeta \ll 1$ and we can expand the function $e^{-|\mu(\theta\zeta)|^2\rho^2}$ appearing in the energy-momentum tensor in $\theta$ before the integration, which simplifies the calculations significantly.\footnote{Due to the Gaussian profile of the beam, large values of $\rho$ do not contribute significantly and $(\theta\zeta)^n\rho^2$ can be considered as small for all $n \geq 1$.} Expressions for $h^\lambda_{\bar\alpha\bar\beta}$ up to first order in $\theta$ for the case of small distances between emitter and absorber of the laser beam can be found in Appendix~\ref{sec:metricpert}. 

In the following, we discuss the metric perturbation in different orders in $\theta$ and present its physical effects. As already the effects in the leading order of our expansion are too small to be measurable with current technology \cite{raetzel_2016}, this will also be the case for the effects in the higher orders. However, the effects are of conceptual interest, as they illustrate the gravitational properties of light.

\section{Zeroth/leading order}
\label{sec:leading}

The metric in the leading order corresponds to the full metric at $\theta=0$, and thus to the metric for the laser  beam in the paraxial approximation. Then, the components of the Poynting vector transversal to the beamline vanish and the only non-zero component of the Maxwell stress tensor is $\sigma^\lambda_{\zeta\zeta}$. Furthermore, $\sigma^\lambda_{\zeta\zeta}=\mathcal{E}^\lambda=-S^\lambda_\zeta/c$, which leads to
\begin{eqnarray}\label{eq:Tmunurel}
	T^{\lambda(0)}_{\bar\alpha\bar\beta} &=&\mathcal{E}^{(0)}\begin{pmatrix}
	1 & 0 & 0 & -1 \\
	0 & 0 & 0 & 0 \\
	0 & 0 & 0 & 0 \\
	-1 & 0 & 0 & 1
	\end{pmatrix} =: \mathcal{E}^{(0)} M^0_{\bar{\alpha}\bar{\beta}}\;,
\end{eqnarray}
where $\mathcal{E}^{(0)} = \varepsilon_0 w_0^2 E_0^2 |v_0|^2 = 2P_0|v_0|^2/(\pi c) $. Therefore, the metric perturbation is found as 
\begin{equation}\label{eq:hfirstorder}
	h^{\lambda(0)}_{\bar\alpha\bar\beta}= I^{(0)} M^0_{\bar{\alpha}\bar{\beta}}\;,
\end{equation} 
where, for the case that the emitter and absorber of the laser beam are far away from each other, we find from Eq.~(\ref{eq:solpoisson}) 
\begin{eqnarray}\label{eq:leadingorderh}
	I^{(0)}
	&=& \frac{\kappa w_0^2 P_0}{2\pi c}  \left(\frac{1}{2}\rm Ei\left(-2|\mu|^2\rho^2\right)-\log(\rho)\right)\;,
\end{eqnarray}
where ${\rm Ei(x)}$ is the exponential integral function. The solution (\ref{eq:leadingorderh}) can be compared with the exact solution derived by Bonnor for an infinitely extended beam of a light-like medium without divergence. The derivation of the metric for a Gaussian profile of the energy density of the medium is given in Appendix \ref{sec:exact}. Bonnor's solution is split into an interior and an exterior solution that are matched at a finite transversal radius $a$. If the beam is infinitely extended in the transverse direction, we are left with an interior solution only which reads
\begin{eqnarray}\label{eq:infbonnor}
	g_{\bar\alpha\bar\beta}^{\rm{B}}
	&=& \eta_{\bar\alpha\bar\beta} - \frac{\kappa w_0^2 P_0}{2\pi c}\left(\log(\rho)-\frac{1}{2}{\rm Ei}\left(-2\rho^2\right)\right)M^0_{\bar\alpha\bar\beta}\;.
\end{eqnarray}
For $\theta =0 $, we have $\mu(\theta\zeta)=1$, and the solution in Eq.~(\ref{eq:leadingorderh}) coincides with (\ref{eq:infbonnor}).

\subsubsection{Small distance between emitter and absorber of the laser beam}

For the case when the emitter and absorber of the laser beam are close to each other, we have to take the second approach described in Sec.~(\ref{sec:metric}). With $\theta\zeta\ll 1$, the retarded potential (\ref{eq:primeli}) in leading order in $\theta$ becomes
\begin{eqnarray}\label{eq:brombeere}
	I^{(0)}
	&=&	\frac{\kappa w_0^2 P_0}{\pi c} e^{-2\rho^2}
		\int_0^\infty d\rho'\; \rho' \log\left(\frac{\beta-\zeta+\sqrt{(\beta-\zeta)^2+\rho'{}^2}}{\alpha-\zeta+\sqrt{(\alpha-\zeta)^2+\rho'{}^2}}\right) J_0\left(i4\rho\rho'\right)
		e^{-2\rho'{}^2}\;,
\end{eqnarray}
where $J_0$ is the Bessel function of the first kind. For small beam waists, $w_0\ll 1$, the solution for the laser  beam (\ref{eq:brombeere}) approaches the solution for the infinitely thin beam (\ref{eq:blaubeere}), as shown in Appendix \ref{sec:infbeam}. We obtain
\begin{eqnarray}\label{eq:blaubeere}
	I^{(0)}_{w_0 \rightarrow 0}
	&=& 	\frac{\kappa w_0^2 P_0}{ 4\pi c}\log\left(\frac{\beta-\zeta+\sqrt{(\beta-\zeta)^2+\rho^2}}{\alpha-\zeta+\sqrt{(\alpha-\zeta)^2+\rho^2}}\right)\;.
\end{eqnarray}
Thus, in the paraxial approximation, we may say that the solution for the laser beam approaches the solution for the infinitely thin beam of constant energy per length of \cite{tolman_1931} as the beam waist goes to zero. Note that the limit $w_0 \rightarrow 0$ can only be considered for the leading order of the laser beam here. This is because $\theta=0$ implies that the condition $\theta \zeta \ll 1$ can be satisfied for all $w_0$. In contrast, for any non-vanishing $\theta$, the conditions $w_0 \rightarrow 0$ and $\theta /w_0 = \theta \zeta \ll 1$ imply $z \rightarrow 0$.

\begin{figure}[H]\center
	\includegraphics[scale=0.4]{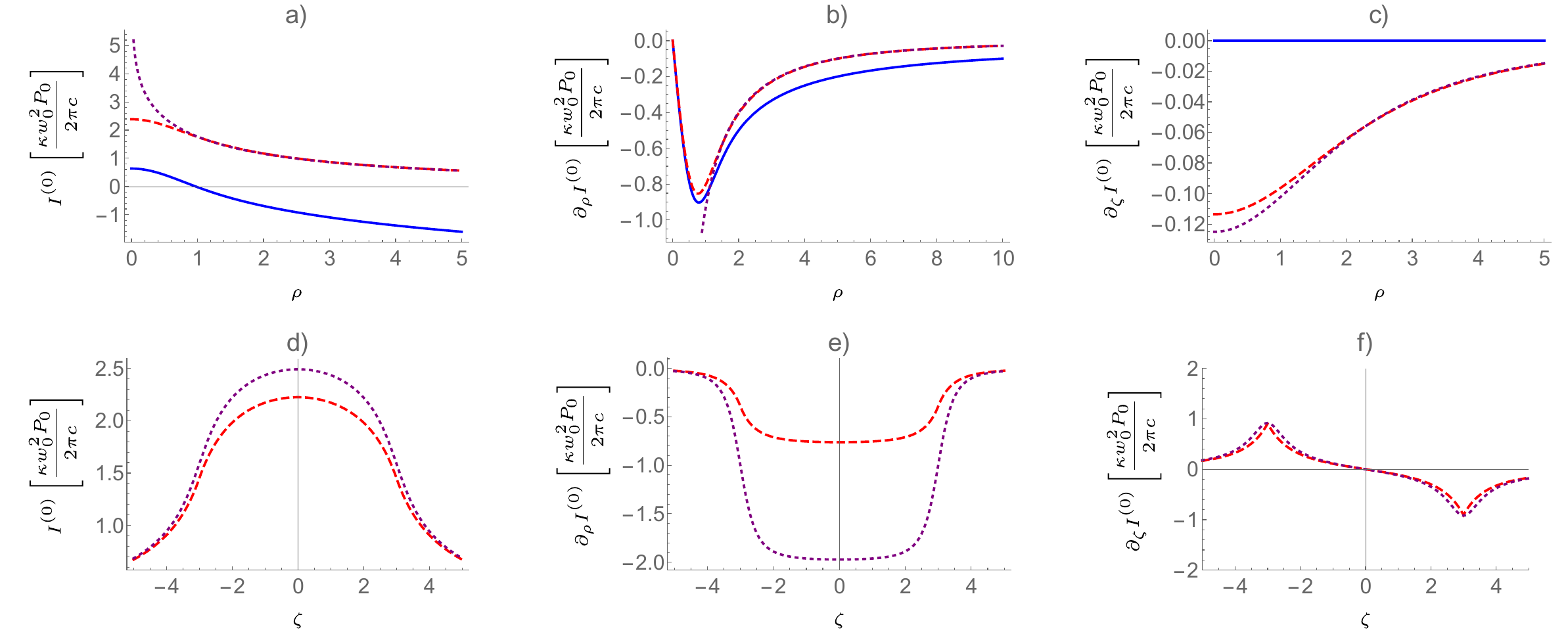}
		\caption{\label{fig:zeroth} These plots show the value of the leading order of the metric perturbation $I^{(0)}$ and its first derivatives for the Gaussian beam with infinite distance between  (plain, blue), the Gaussian beam with short distance between emitter and absorber of the laser beam (dashed, red), and the infinitely thin beam (dotted, purple) in units of $\kappa P_0 w_0^2/(2\pi c)$. In the second and the third cases, the distance between laser beam's emitter and absorber is chosen to be $6$. In the first row, the functions are plotted for $\zeta=1$ and in the second row for $\rho=1/2$. The second row does not contain plots for the case of large distances between emitter and absorber of the laser beam as there is no dependence of $I^{(0)}$ on $\zeta$ in that case. {We find that the values for $I^{(0)}$ and its first derivatives are usually larger for the infinitely thin beam than for the other two cases. This is due to the divergence at the beamline for the case of the infinitely thin beam. In the other two cases, the gravitational field is spread out as the sources are.} In b), we see that the absolute value of the first $\rho$-derivative of $I^{(0)}$ reaches a maximum at a finite distance from the beamline. Note that $\partial_\rho I^{(0)}$ is proportional to the acceleration that a test particle experiences if it is initially at rest at a given distance $\rho$ to the beamline.  We see that acceleration is always directed towards the beamline. It is larger in the case of an infinite distance between emitter and absorber of the laser beam than in the case of a finite distance, which we can attribute to the larger extension of the source (and thus the larger amount of energy) in the former than in the latter. In e), which shows plots for the cases of finite distance between the laser beam's emitter and absorber, we see that $\partial_\rho I^{(0)}$ still is the largest at the center between emitter and absorber of the laser beam and decays quickly once their positions at $\zeta=\pm 3$ are passed. $\partial_\zeta I^{(0)}$ is proportional to the acceleration in the $\zeta$-direction. As expected it vanishes for infinite distance between emitter and absorber of the laser beam. In f), we see that the acceleration is directed towards the center between laser beam's emitter and absorber and its absolute values reaches its maximum at $\zeta=-3$ and $\zeta=3$, the $\zeta$-coordinates of emitter and absorber of the laser beam respectively. }
\end{figure}

In Fig.~\ref{fig:zeroth}, the function $I^{(0)}$ and its derivatives are illustrated for the the infinitely long Gaussian beam, the Gaussian beam with short distance between emitter and absorber, and the infinitely thin beam. 

\subsection{Acceleration of a test particle at rest}

Let us consider the acceleration a massive test particle would experience if it was initially at rest 
at given $\rho$ and $\zeta$. Then, the initial normalized tangent to its worldline $\gamma(\tilde\tau)$, where $\tilde\tau$ is the proper time, is given as $\dot\gamma=c w_0^{-1}\big(1+h_{\tau\tau}^{\lambda(0)}/(2w_0^2),0,0,0\big)$, where the dot refers to the derivative with respect to proper time. From the geodesic equation (\ref{apfel}) and the form of the metric in zeroth order, we find that 
\begin{eqnarray}
	\ddot\gamma^\rho\simeq 
	\frac{c^2}{2w_0^4}\partial_\rho I^{(0)}
	\quad\text{and}\quad 
	\ddot\gamma^\zeta
	\simeq \frac{c^2}{2w_0^4}\partial_\zeta I^{(0)}	
	\;.
\end{eqnarray}
Plots of $\partial_\rho I^{(0)}$ and $\partial_\zeta I^{(0)}$ for the three different cases above are given in Fig.~\ref{fig:zeroth}. As a numerical example for the long beam, for the power $P_0\sim 10^{15}{\rm W}$, the beam waist $w_0\sim 10^{-3}{\rm m}$, a particle at rest at the location $z=0$ and $r=\sqrt{x^2+y^2}=w_0$ is accelerated by $\ddot\gamma^r\sim -10^{-18}{\rm ms^{-2}}$. \footnote{Here and in the following numerical examples, in order to express the acceleration in the coordinates $\{ct,x,y,z\}$, the Leibnitz rule has been applied and it has been used that the difference between proper time and coordinate time is proportional to the metric perturbation.} This is of the same order of magnitude as for the infinitely thin beam \cite{raetzel_2016}.

\subsection{Curvature}

For the leading order, we can find the components of the curvature tensor using Eq.~(\ref{eq:Rexprelation}) in Appendix \ref{sec:Rexpansion} and Eq.~(\ref{eq:hfirstorder}). 
The only non-zero independent components of the Riemann curvature tensor for the metric perturbation given in Eq.~(\ref{eq:leadingorderh}) and Eq.~(\ref{eq:brombeere}) and the limit of an infinitely thin beam in Eq.~(\ref{eq:blaubeere}) are 
\begin{eqnarray}\label{eq:riemleading}
	 r_{\tau \bar i\tau\bar j}^{(0)}= r^{(0)}_{\zeta\bar i\zeta \bar j}= -r_{\tau\bar i\zeta\bar j}^{(0)}
	&=&	-\frac{1}{2}\partial_{\bar i}\partial_{\bar j} I^{(0)}\;.
\end{eqnarray}
For the case of a far extended beam neglecting emitter and absorber of the laser beam that was given in Eq.~(\ref{eq:leadingorderh}), we obtain 
\begin{eqnarray}
	 r^{(0)}_{\tau\xi\tau\xi}= r^{(0)}_{\zeta\xi\zeta\xi}= -r^{(0)}_{\tau\xi\zeta\xi}
	&=& -\frac{\kappa w_0^2 P_0}{4\pi c}\,\frac{1}{\rho^4}\left((\xi^2-\chi^2)-\left(4 |\mu|^2\xi^2\rho^2+\xi^2-\chi^2 \right) e^{-2|\mu|^2\rho^2}\right)
	\;,\\
	r^{(0)}_{\tau\chi\tau\chi}= r^{(0)}_{\zeta\chi\zeta\chi}= -r^{(0)}_{\tau\chi\zeta\chi}
	&=& \frac{\kappa w_0^2 P_0}{4\pi c}\,\frac{1}{\rho^4}\left((\xi^2-\chi^2)+\left(4 |\mu|^2\chi^2\rho^2-\xi^2+\chi^2 \right)e^{-2|\mu|^2\rho^2}\right)
	\;,\\
	r^{(0)}_{\tau\xi\tau\chi}= r^{(0)}_{\zeta\xi\zeta\chi}= -r^{(0)}_{\tau\xi\zeta\chi}
	&=& -\frac{\kappa w_0^2 P_0}{2\pi c}\,\frac{\xi\chi}{\rho^4}\left(1-(1+2|\mu|^2 \rho^2) e^{-2|\mu|^2\rho^2}\right)
	\;.
\end{eqnarray} 

\subsubsection{Comparison to the infinitely thin beam}

In the paraxial approximation (i.e.~for $\theta=0$) and for small beam waists, the Riemann curvature tensor of the infinitely long laser  beam approaches the Riemann curvature tensor of the infinitely thin beam, as does the metric. It is also interesting to compare the curvature for the infinitely thin beam with that for the full solution given in \cite{bonnor_1969} by Bonnor. The analysis can be found in Appendix \ref{sec:exact} for a beam with a Gaussian profile cut off at a radius $a$. The corresponding solution splits into an interior solution and an exterior solution. For $a\rightarrow \infty$, we obtain the solution in Eq.~(\ref{eq:infbonnor}) that we compared with our leading order metric perturbation already. In Appendix \ref{sec:exact}, we give the components of the curvature tensor in the exterior region ($r>a$) in Eq.~(\ref{eq:extcurvbonnor}). We show that it coincides with the components of the curvature tensor of an infinitely thin beam. In particular, the curvature is independent of the radial dependence of the beam intensity; only the total power of the beam is important. 

\section{First order and frame dragging}
\label{sec:firstorderframedrag}

The metric perturbation for large distances between emitter and absorber of the laser beam 
in first order in $\theta$ is determined by the first order of the energy-momentum tensor, $\bar{t}^{\lambda(1)}_{\bar\alpha\bar\beta}$, which has the only independent non-zero components
\begin{eqnarray}\label{jup}
	\theta \bar{t}^{\lambda(1)}_{\tau\xi} = -\theta \bar{t}^{\lambda(1)}_{\zeta\xi} = -S_\xi^{\lambda(1)} /c
	&=&	- \mathcal{E}^{(0)} \theta |\mu|^2 (\theta \zeta  \xi +\lambda  \chi )\;,\\\A
	\theta \bar{t}^{\lambda(1)}_{\tau\chi} = -\theta \bar{t}^{\lambda(1)}_{\zeta\chi} = -S_\chi^{\lambda(1)} / c
	&=&	\lambda \mathcal{E}^{(0)} \theta |\mu|^2  (  \xi - \lambda \theta\zeta \chi )\;.
\end{eqnarray}
Note that $\tilde\zeta=\theta\zeta$ is the coordinate that is considered for the asymptotic expansion in Eqs.~(\ref{eq:parhelm}), (\ref{eq:parhelm1}) and (\ref{eq:parhelmhigher}).
Therefore, $S_\xi^{\lambda(1)}$ and $S_\chi^{\lambda(1)}$ are indeed of first order in $\theta$ regarding the expansion (\ref{eq:expansionv}). 

From Eq.~(\ref{eq:poissonh1}), we obtain for the metric perturbation in first order in $\theta$
\begin{eqnarray}
	h^{\lambda(1)}_{\bar\alpha\bar\beta}
	&=&	\begin{pmatrix}
			0& I_\xi^{\lambda(1)} & I_\chi^{\lambda(1)} & 0\\
			I_\xi^{\lambda(1)} & 0 & 0 & -I_\xi^{\lambda(1)}\\
			I_\chi^{\lambda(1)} & 0 & 0 & -I_\chi^{\lambda(1)}\\
			0 & -I_\xi^{\lambda(1)} & -I_\chi^{\lambda(1)} & 0
		\end{pmatrix}\;,
\end{eqnarray}
where 
\begin{eqnarray}\label{jup}
	I_\xi^{\lambda(1)}
	&=&	\frac{1}{4}(\theta \zeta  \partial_\xi + \lambda  \partial_\chi )I^{(0)}= -\frac{\kappa  P_0 w_0^2 (\theta \zeta  \xi +\lambda  \chi )}{8 \pi  c \rho^2}  \left(1-e^{-2 |\mu|^2\rho^2}\right)\;,\\
	I_\chi^{\lambda(1)}
	&=&	- \frac{1}{4}( \lambda\partial_\xi - \theta\zeta \partial_\chi )I^{(0)} = \frac{\kappa  P_0 w_0^2 (\lambda  \xi - \theta \zeta  \chi )}{8 \pi  c \rho^2}  \left(1-e^{-2 |\mu|^2\rho^2}\right)\;.
\end{eqnarray}
For small $\theta\zeta$, the terms proportional to $\theta\zeta$ can be neglected in (\ref{jup}) such that we find
\begin{eqnarray}\label{jupsmall}
	I_\xi^{\lambda(1)}
	&=&	\frac{\lambda}{4}\partial_\chi I^{(0)}\quad {\rm and}\quad
	I_\chi^{\lambda(1)}
	=	- \frac{\lambda}{4}\partial_\xi I^{(0)}\;.
\end{eqnarray}
It is interesting to note that our solution coincides with the exact solution of Einstein's equations presented in \cite{bonnor_1970} by Bonnor for a rotating null fluid. In particular, we can identify our functions in the metric with those of \cite{bonnor_1970} as $\alpha= \theta I^{\lambda(1)}_\chi/\sqrt{2}$, $\beta =  \theta I_\xi^{\lambda(1)}/\sqrt{2}$ and $A=I^{(0)}$. Our equation (\ref{eq:poissonh1}) corresponds to the equations (2.16) and (2.17) in \cite{bonnor_1970}. Similar expressions for the metric of a circularly polarized light beam are presented in \cite{frolov_2005}.

\subsubsection{Small distance between emitter and absorber of the laser beam}

\begin{figure}[H]\center
	\includegraphics[scale=0.37]{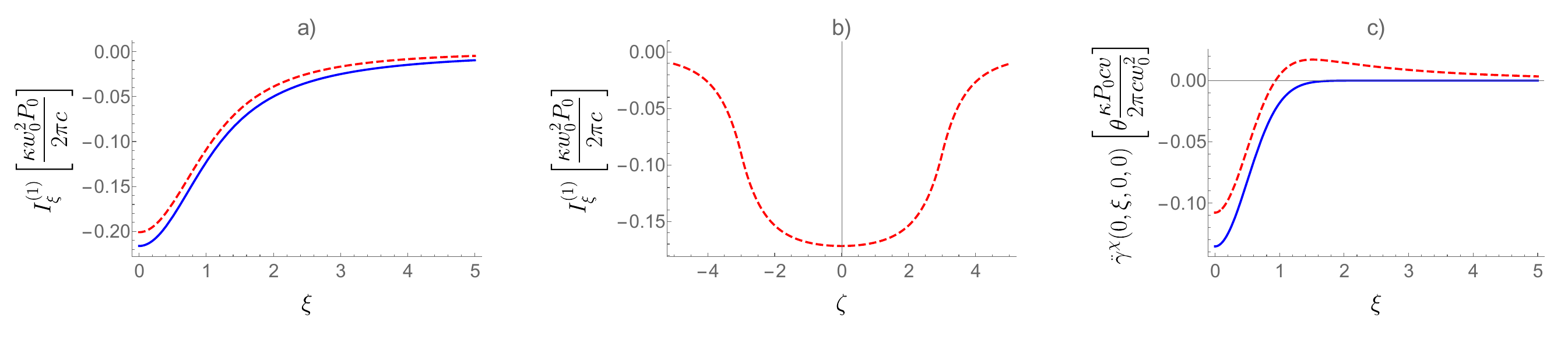}
	\caption{\label{fig:firstorder} Considering $\theta\zeta\ll 1$, the first two plots show the function $I^{(1)}_\xi$ for an infinite distance between emitter and absorber of the laser beam (plain, blue) and a short distance between laser beam's emitter and absorber (dashed, red) as a function of $\xi$ for $\zeta=0.1$ and $\chi=0$ (plot a) and as a function of $\zeta$ at $\xi=1/2$ and $\chi=1$ (plot b). The functions are plotted in units of $\kappa w_0^2 P_0/(2\pi c)$. In b) there is no plot for the case of infinite distance between emitter and absorber of the laser beam as the result does not depend on $\theta\zeta$. Plot c) shows the deflection in the $\chi$-direction a light test particle would experience if it would move radially outwards in the $\xi$-direction at $\chi=0$ for an infinite distance between emitter and absorber of the laser beam (plain, blue) and a short distance between laser beam's emitter and absorber (dashed, red). This effect is induced by frame dragging. We see that the effect changes sign for the case of a short distance between laser beam's emitter and absorber. }
\end{figure}

For small distances between emitter and absorber of the laser beam, we find directly Eq.~(\ref{jupsmall}), where $I^{(0)}$ has to be taken from Eq.~(\ref{eq:brombeere}). In Fig.~\ref{fig:firstorder}, the function $I_\xi^{\lambda(1)}$ is illustrated as a function of $\xi$ and $\zeta$ for $\chi=1$. The plots for $I_\chi^{\lambda(1)}$ would look similar when plotted as a function of $\chi$ and $\zeta$ for $\xi=1$.

\subsubsection{Curvature}

It was shown in \cite{bonnor_1970} that the rotation of the null fluid leads to frame dragging. This has been shown to be the case as well in \cite{strohaber_2013} for a laser  beam of light with angular momentum. Here, we obtain the frame dragging effect in the curvature tensor components. The only non-zero components of first order (see Eq.~(\ref{eq:Rexprelation})) are
\begin{eqnarray}
		\A r^{\lambda(1)}_{\bar j\zeta \bar j\bar k} &=&  -\frac{1}{2}\partial_{\bar j}\left(\partial_{\bar j} h^{\lambda(1)}_{\zeta\bar k} - \partial_{\bar k} h^{\lambda(1)}_{\zeta \bar j}\right)\;,\\
		 r^{\lambda(1)}_{\bar j \tau \bar j\bar k} &=& - \frac{1}{2}\partial_{\bar j}\left(\partial_{\bar j} h^{\lambda(1)}_{\tau\bar k} - \partial_{\bar k} h^{\lambda(1)}_{\tau \bar j}\right)\;,\\
		\A r^{\lambda(1)}_{\bar j \tau \zeta \tau} &=&  -\frac{1}{2}\partial_{\bar j}\partial_{\theta\zeta} h^{\lambda(0)}_{\tau\tau} \;,
\end{eqnarray}
where $\bar j\neq \bar k$. For small $\theta\zeta$, we can neglect $r^{\lambda(1)}_{\bar j \tau \zeta \tau}$ and we find 
\begin{eqnarray}\label{eq:curvfirst}
		 r^{\lambda(1)}_{\xi \zeta \xi \chi} = -\lambda\frac{\kappa w_0^2}{2}\xi \mathcal{E}^{(0)} = -r^{\lambda(1)}_{\xi \tau \xi \chi} \quad\mathrm{and}\quad 
		 r^{\lambda(1)}_{\chi \zeta \chi \xi} = \lambda\frac{\kappa w_0^2}{2}\chi \mathcal{E}^{(0)} = -r^{\lambda(1)}_{\chi \tau \chi \xi}\;.
\end{eqnarray}
The non-zero curvature components $r^{\lambda(1)}_{\xi \tau \xi \chi}$ and $r^{\lambda(1)}_{\chi \tau \chi \xi}$ lead to the precession of gyroscopes, which can be seen most straight forward in the framework of gravitomagnetism \cite{Mashhoon:2003ax}; they can be interpreted as gravitomagnetic fields that govern the motion of test particles in a gravitational Lorentz force law. 

\subsubsection{Deflection of test particles}

The frame dragging effect can be studied alternatively using the geodesic equation (\ref{apfel}) and the expressions for the Christoffel symbols in Eq.~(\ref{eq:gammaexprelation}). Let us consider a test particle moving radially outwards with velocity $v$. We will only consider terms linear in $v/c$ in the following. Then, the initial tangent $\dot\gamma(0)=cw_0^{-1}\big( 1-f,v/c,0,0 \big)$ to the test particle's world line $\gamma(\tilde \tau)$ at $\gamma(0)=(0,\xi,0,0)$ and $f(v,\xi,\chi,\theta\zeta)$ is chosen such that $\dot\gamma(\tilde\tau)$ fulfills the condition $g_{\bar\mu\bar\nu}(\gamma(\bar\tau))\dot\gamma^{\bar\mu}(\bar\tau)\dot\gamma^{\bar\nu}(\bar\tau)=-c^2$ at $\tilde\tau=0$, where again $\tilde\tau$ is the proper time and the dot represents the derivative with respect to it.
In first order in the metric perturbation, we find that
\begin{equation}\label{eq:drag}
	\ddot\gamma^\chi (0) = \frac{c v}{w_0^4} \theta \left(\partial_\chi h_{\tau\xi}^{(1)} - \partial_\xi h_{\tau\chi}^{(1)} \right) = -\lambda v\,\theta\frac{\kappa P_0}{2\pi w_0^2}|\mu|^2e^{-2 |\mu|^2\rho^2}\;.
\end{equation} 
We see that massive test particles do not propagate radially. Their trajectories are transversally bent, where the sign of the bending depends on the polarization of the laser  beam. This is the effect of frame dragging. 
For $v\sim 10{\rm ms^{-1}}$, $P_0\sim 10^{15}{\rm W}$, $\theta\sim 10^{-3}$, $w_0\sim 10^{-3}{\rm m}$, $z=0$ and $x=w_0$, the acceleration is of the order of magnitude ${d^2\gamma^y(0)/dt^2}\sim \pm 10^{-29}{\rm ms^{-2}}$.

\begin{figure}[H]\center
	\includegraphics[scale=0.4]{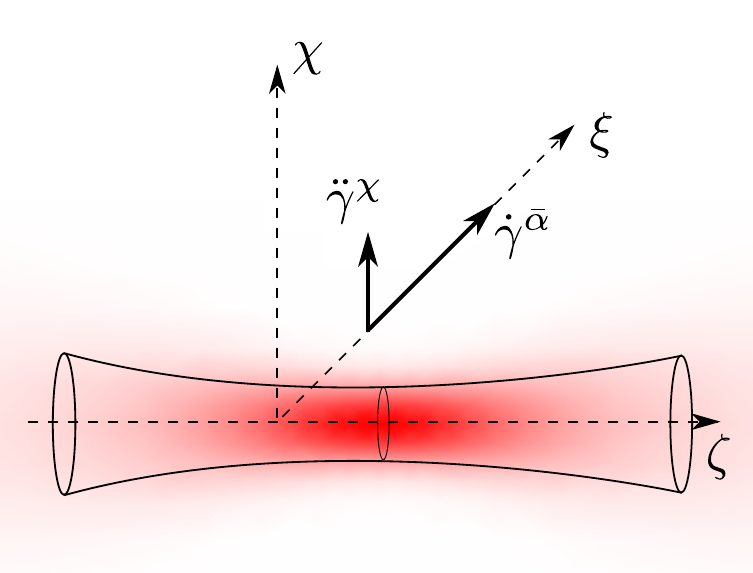}
	\caption{\label{fig:fd} Schematic illustration of the frame dragging effect: A massive particle moving radially outwards from the beamline (here in $\xi$-direction) is accelerated in the transverse direction (here in $\chi$-direction).}
\end{figure}

The effect in Eq.~(\ref{eq:drag}) decreases exponentially with the distance to the beamline.
The same is true for the curvature components in Eq.~(\ref{eq:curvfirst}). 
The effect is due to the spin angular momentum due to the helicity of the beam. In contrast, in \cite{strohaber_2013}, frame dragging effects for $\rho \gg 1$ have been shown 
to arise from the orbital angular momentum of optical vortices. In Fig.~\ref{fig:fd}, the above deflection is illustrated.
It is interesting to note that, by direct calculation from the expressions for the metric perturbation up to third order in $\theta$ in Appendix \ref{sec:mpert}, we find for $\rho\gg 1$
\begin{equation}
	R_{\bar\alpha\bar\beta\bar\gamma\bar\delta}\approx \left(1+\frac{\theta^2}{2}\right) R^{(0)}_{\bar\alpha\bar\beta\bar\gamma\bar\delta}\,,
\end{equation}
up to third order in $\theta$. All other terms decay exponentially with $\rho^2$. Therefore, far away from the beam and up to third order in $\theta$, there are no effects beyond those that already exist in zeroth order. All additional effects appear only where the energy distribution of the source beam is non-vanishing; they are effects of a local gravitational coupling between the source and test particles. In the next section, we will discuss another such effect in fourth order in $\theta$, the deflection of parallel co-propagating test rays.

\section{Fourth order - The deflection of parallel co-propagating test rays}
\label{sec:fourthorderdefl}

\begin{figure}[H]\center
	\includegraphics[scale=0.4]{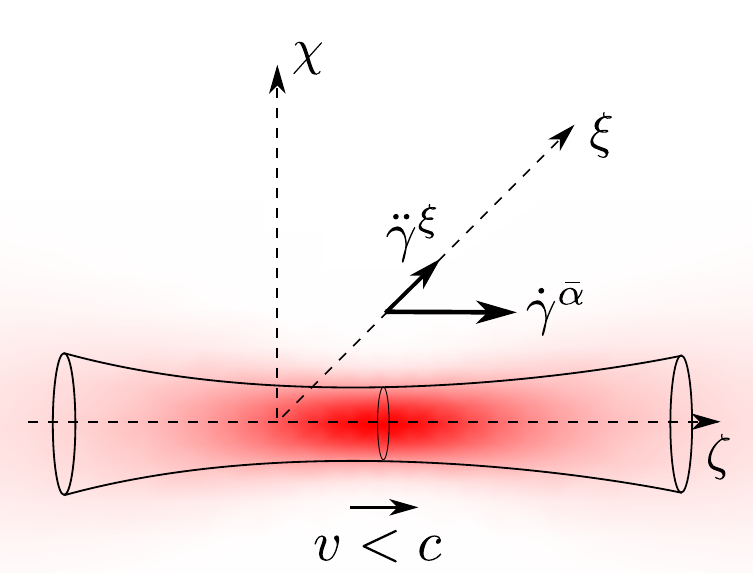}
	\caption{\label{fig:beam} Schematic illustration of the source laser beam and the parallel co-propagating test ray of light: We look at the deflection of the test ray of light due to the gravitational field of the laser beam.}
\end{figure}

As discussed in \cite{tolman_1931} for a finitely long and infinitely thin light beam, a test ray propagating parallel to it is not deflected. It has also been shown \cite{bonnor_1969} that the superposition of two exact solutions of the Einstein equations for pp-waves travelling in the same direction is again a solution, confirming the result of the linearized theory. 
In our description, there are two more important characteristics of the laser beam playing an important role, both of them coming from the wave-like nature of light: First, the laser beam is diverging. Second, in \cite{federov_2017}, it was argued that light in a laser beam does not move with the speed of light along the beamline, but with a slightly smaller velocity. The origin of the effect is the superposition of plane waves with different wave vectors, which leads to a reduced effective propagation speed. This was confirmed experimentally in \cite{Giovannini:2015spa}. In \cite{federov_2017}, the difference between the speed of light and the group velocity of light in a laser beam was found to be\footnote{In \cite{federov_2017}, a different definition of the beam waist is used (see Eq.~(28) of \cite{federov_2017}) such that $w = w_0/\sqrt{2}$ in Eq.~(40) of \cite{federov_2017}.} $\delta v_\theta = c - v_\theta = c/(k^2w_0^2) = c\theta^2/4$.
It has been shown by Scully that two parallel co-propagating thin beams in a wave-guide, since they are propagating slower than the speed of light, do gravitationally interact with each other \cite{scully_1979}. Therefore one may wonder whether the source laser beam deflects an originally parallel co-propagating test ray. We will investigate this question in the following.
The setup is illustrated in Fig.~\ref{fig:beam}.

A parallel co-propagating test light-ray is described by the light-like tangent vector $\dot\gamma^{\bar\alpha}= w_0^{-1}c(1,0,0,1-f)$, where $f$ is determined by the null-condition and found to be of the same order of magnitude as the metric perturbation, and therefore does not contribute in the following, and again the curve is parametrized with proper time and the dot stands for the derivative with respect to it. 
With the geodesic equation, we obtain
\begin{eqnarray}
	\ddot\gamma_+^{\bar j} &=& -c^2 w_0^{-2} \left[ \Gamma^{\bar j}_{\tau\tau} + 2\Gamma^{\bar j}_{\tau\zeta} + \Gamma^{\bar j}_{\zeta\zeta}\right] \\\label{blume}
	&=& -c^2 w_0^{-4} \theta^4\left[ -\frac{1}{2}\partial_{\bar j}\left(h_{\tau\tau}^{(4)} + 2 h_{\tau\zeta}^{(4)} + h_{\zeta\zeta}^{(4)}\right) + \partial_{\theta\zeta}\left( h_{\tau \bar j}^{(3)} + h_{\zeta \bar j}^{(3)}\right)\right]\;.
\end{eqnarray}
From the expression for the components of the energy-momentum tensor in Appendix \ref{sec:pointing} and Eqs.~(\ref{eq:poissonhhigher}), we find 
\begin{eqnarray}
	\Delta_\mathrm{2d} \left(h_{\tau\tau}^{(4)} + 2 h_{\tau\zeta}^{(4)} + h_{\zeta\zeta}^{(4)} \right) &=& -\frac{\kappa w_0^2 P_0}{2\pi c}|\mu|^6\rho^4  e^{-2|\mu|^2\rho^2}\;, 
\end{eqnarray}
which is solved by Eq.~(\ref{eq:solpoisson}) as
\begin{eqnarray}
	h_{\tau\tau}^{(4)} + 2 h_{\tau\zeta}^{(4)} + h_{\zeta\zeta}^{(4)} &=& \frac{\kappa w_0^2 P_0}{32\pi c} \Bigg(\text{Ei}\left(-2 |\mu|^2\rho^2 \right)-2 \log (\rho) - \left(\frac{3}{2} + |\mu|^2\rho^2\right) e^{-2|\mu|^2\rho^2} \Bigg)\;.
\end{eqnarray}
The components of the metric perturbation in third order in $\theta$ which appear in the second term in equation (\ref{blume}) can be found in Appendix \ref{sec:mpert}.
We obtain for $\bar j\in \{\xi,\chi\}$, assuming that $\theta\zeta\ll 1$,
\begin{eqnarray}\label{eq:deflfourth}
	\ddot\gamma^{\bar j} = \frac{c\kappa P_0}{32 \pi w_0^2}\theta^4 \frac{x^{\bar j}}{\rho^2} \left(1-(1 - 2\rho^4) e^{-2\rho^2} \right)\,.
\end{eqnarray}
For large distances from the beamline ($\rho \gg 1$) and $j\in\{x,y\}$, the acceleration becomes
\begin{eqnarray}\label{eq:coordacc}
	\ddot\gamma^{\bar j} = \frac{c\kappa P_0 }{32 \pi w_0^2}\theta^4 \frac{x^{ \bar j}}{\rho^2} \;,
\end{eqnarray}
which is an apparent repulsion. This is due to the second term in Eq.~(\ref{blume}). If we had considered only the first term in Eq.~(\ref{blume}), we would have obtained the same absolute acceleration as in Eq.~(\ref{eq:coordacc}), but with the opposite sign. Hence, the first term in Eq.~(\ref{blume}) induces an attraction and the second term a repulsion.

However, coordinate acceleration does not have any physical meaning in general relativity. Therefore, we have to investigate the geodesic deviation to learn about the meaning of the coordinate acceleration (\ref{eq:coordacc}). With the separation vector $s^{\bar\alpha}=(0,1,0,0)$ and the tangent $\dot\gamma^{\bar\alpha}= w_0^{-1}c(1,0,0,1-f)$, we obtain for the acceleration of the separation vector in $\xi$-direction from Eq.~(\ref{eq:geodesicdev})
\begin{equation}\label{eq:properaccfourth}
	a^\xi
	=\frac{\theta^4c^2}{2w_0^4}\left(
	\partial_\xi^2\left(h_{\tau\tau}^{(4)}+ 2h_{\tau\zeta}^{(4)}+h_{\zeta\zeta}^{(4)}\right)
	-2\partial_\xi\partial_{\theta \zeta}\left( h_{\tau\xi}^{(3)}+h_{\xi\zeta}^{(3)}
	\right)
	+\partial_{\theta \zeta}^2h_{\xi\xi}^{(2)}\right)\;.
\end{equation}
With the expressions for the combinations of the metric perturbation given above and in appendix F, we obtain in the case of $\theta\zeta\ll 1$
\begin{eqnarray}\label{ameise}
	a^\xi
	&=&	-\frac{\kappa c P_0 \theta^4}{16\pi w_0^2 }e^{-2 \rho^2}
	\left( \rho^2(4\xi^2+3)-6\xi^2\right)\;,
\end{eqnarray}
which vanishes far from the beamline. Therefore, we found that the deflection in Eq.~(\ref{eq:deflfourth}) is a coordinate effect. More precisely, the geodesic deviation in Eq.~(\ref{eq:properaccfourth}) can be split into two parts. The first part is the $\xi$-derivative of the coordinate acceleration $\ddot\gamma^{\bar j}$ in Eq.~(\ref{eq:deflfourth}). The second part is the second $\theta\zeta$-derivative of $h_{\zeta\zeta}^{(2)}$ which corresponds to the change of the definition of length in the $\xi$-direction. The contributions of the two parts to the geodesic deviation cancel for large distances from the beamline.  
As a numerical example, for $P_0\sim 10^{15}{\rm W}$, $\theta\sim 10^{-3}$, $w_0\sim 10^{-3}{\rm m}$, $x=w_0$ and $y=0$, one has
$a^x\sim -10^{-31}{\rm ms^{-2}}$. Notice that this is the relative acceleration of two test light-rays. The interesting point is that it is non-zero.

\subsection{Comparison to the boosted infinitely long massive cylinder} 

\begin{figure}[H]\center
	\includegraphics[scale=0.4]{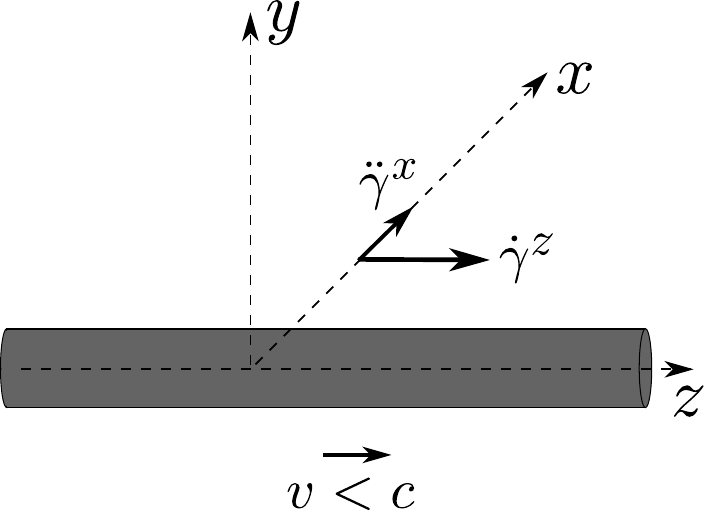}
	\caption{\label{fig:rod} A massive cylinder moving at the speed $v<c$ and a parallel co-propagating test light-beam: We investigate the gravitational deflection of the test beam due to the gravitational field of the cylinder.}
\end{figure}

The reduced propagation speed argued for in \cite{federov_2017} suggests that the result in Eq.~(\ref{eq:deflfourth}) may be compared to the deflection of a parallel test ray by a cylindrically symmetric mass distribution moving with $v=c-\delta v_\theta$ along the cylinder axis (see Fig.~\ref{fig:rod}). That is the content of this subsection.

The exterior gravitational field of a cylindrically symmetric mass distribution of rest of mass per unit length $\mathfrak{m}$ (dimensionless units) is described by the Levi-Civita metric (originally presented in \cite{levicivita_1919}; it can be found in \cite{bonnor_1992} containing four additional parameters $K$, $L$, $M$ and $N$ that we set equal to one),
\begin{eqnarray}
	ds^2
	&=&	-\rho^{4{\mathfrak{m}}}c^2 dt^2+\rho^{8{\mathfrak{m}}^2
	-4{\mathfrak{m}}}\left(d\rho^2+dz^2\right)
	+\rho^{2-4{\mathfrak{m}}}d\phi^2\;,
\end{eqnarray}
in the cylindrical coordinates $(ct,\rho,\phi,z)$, where $\rho=\sqrt{x^2+y^2}$. The parameter ${\mathfrak{m}}$ can be considered to be a dimensionless quantity representing the mass or energy per unit length for $0<{\mathfrak{m}}<\frac{1}{2}$ \cite{bonnor_1992}. 
Now, we let the cylinder move in positive $\zeta$-direction with normalized velocity $\beta = v/c$, and thus make the coordinate transformation
\begin{eqnarray}
	ct	&\rightarrow&\gamma(ct-\beta z)\;,\\\A
	z 	&\rightarrow& \gamma(z-\beta ct)\;,
\end{eqnarray}
where $\gamma=(1-\beta^2)^{-1/2}$ and $\beta=v/c$. The line density of energy $\mathfrak{m}$ is a quotient of an energy scale $\mathcal{E}$ and a length scale $L$. The energy seen by an observer in the rest frame is $\mathcal{E}'=\gamma \mathcal{E}$. Due to Lorentz contraction, the length scale seen in the rest frame becomes $L'=L/\gamma$. Therefore, the line density of energy seen in the rest frame becomes $\mathfrak{m}'=\gamma^2\mathfrak{m}$. Then, the metric becomes
\begin{eqnarray}
	ds^2
	&=&	\gamma^2\left(-\rho^{4\gamma^{-2}\mathfrak{m}^\prime}+\beta^2\rho^{8\gamma^{-4}\mathfrak{m}^{\prime 2}-4\gamma^{-2}\mathfrak{m}^\prime}\right)
		c^2dt^2
		-2\gamma^2 \beta \left(-\rho^{4\gamma^{-2}\mathfrak{m}^\prime}+\rho^{8\gamma^{-4}\mathfrak{m}^{\prime 2}-4\gamma^{-2}\mathfrak{m}^\prime}\right)
		cdt dz\\\A
	&&	+\gamma^2\left(-\beta^2\rho^{4\gamma^{-2}\mathfrak{m}^\prime}+\rho^{8\gamma^{-4}\mathfrak{m}^{\prime 2} - 4\gamma^{-2}\mathfrak{m}^\prime}\right)
		dz^2
		+\rho^{8\gamma^{-4}\mathfrak{m}^{\prime 2} - 4\gamma^{-2}\mathfrak{m}^\prime}d\rho^2
		+\rho^{2-4\gamma^{-2}\mathfrak{m}^\prime}d\phi^2\;.
\end{eqnarray}
Transforming to cylindrical coordinates according to $\rho=\sqrt{x^2+y^2}$ and $d\rho=\frac{1}{\rho}\left(x dx+y dy\right)$, as well as $d\phi=\frac{1}{\rho^2}\left(-y dx+x dy\right)$, 
and assuming $\gamma^{-2}\mathfrak{m}^{\prime}$ to be small and expanding the terms $\rho^{\gamma^{-2}\mathfrak{m}^\prime}$ as $\rho^\mathfrak{m}=1+\gamma^{-2}\mathfrak{m}^\prime\log(\rho)$ and neglecting quadratic terms in $\gamma^{-2}\mathfrak{m}^\prime$, we obtain
\begin{eqnarray}
	ds^2
	&=& -\left(1+(1+\beta^2)4\mathfrak{m}^\prime\log(\rho)\right)c^2 dt^2
		+16\beta\mathfrak{m}'\log(\rho)cdt dz
		+\left(1-(1+\beta^2)4\mathfrak{m}^\prime\log(\rho)\right)dz^2\\\A
	&&	+\left(1-(1-\beta^2)4\mathfrak{m}^\prime\log(\rho)\right)\left(dx^2+dy^2\right)\;.	
\end{eqnarray}
This metric can be decomposed into the Minkowski metric plus the small perturbation
\begin{eqnarray}\label{uv}
	h_{\alpha\beta}
	&=&	-4\mathfrak{m}^\prime\log(\rho)
		\begin{pmatrix}
			1+\beta^2 &0&0&-2\beta\\
			0&1-\beta^2&0&0\\
			0&0&1-\beta^2&0\\
			-2\beta &0&0&1+\beta^2
		\end{pmatrix}\;.
\end{eqnarray}
We can identify the line density of energy with that of a beam of light as $\mathfrak{m}^\prime c^4/G = P_0/c$. Then, the metric $\eta_{\alpha\beta}+h_{\alpha\beta}$ coincides with the metric of an infinitely long beam of light with constant energy density $P_0/(\pi w_0^2 c)$ confined to a cross section of $\pi w_0^2$ for $\beta=1$ given in \cite{bonnor_1969}, up to constants.

From the metric (\ref{uv}), we find that the parallel test ray with tangent $\dot\gamma^\mu=cL(1,0,0,1)$ is deflected in $x$-direction according to, where again the dots stand for the derivative with respect to proper time,
\begin{eqnarray}\label{eq:deflcyll}
	\ddot\gamma^x
	&=& -\frac{4GP_0}{c^3}\frac{x}{\rho^2}(1-\beta)^2 \;.
\end{eqnarray}
Assuming $1-\beta=\delta v/c = \theta^2/4$, we find that the result in Eq.~(\ref{eq:deflcyll}) differs from Eq.~(\ref{eq:coordacc}) by its sign and a factor $1/2$. Considering the geodesic deviation with the separation vector $s^{\alpha}= dL(0,1,0,0)$, where $dL$ is an infinitesimal length, we obtain
\begin{equation} 
a^x
= \frac{c^2dL}{2}\partial_x^2\left(\frac{1}{c^2}h_{tt}+{\frac{2}{c}}h_{tz}+h_{zz}\right)\;,
\end{equation}
and, inserting the expressions for the metric,
\begin{equation}\label{eq:geodevrod}
	a^x
	= \frac{4GP_0dL}{c^3}  \frac{x^2-y^2}{(x^2+y^2)^2}(1-\beta)^2\;.
\end{equation}
In contrast, for the gravitational field of the focused laser beam, we did not find a deflection for large distances. From this result we see that the gravitational field of light in a Gaussian beam does not 
simply behave as massive matter moving with the velocity derived in \cite{federov_2017} along the beamline. The reason is that the divergence of the laser beam does not only lead to a reduced group velocity, but also to a change of the metric along the beamline. This leads to the appearance of the second and third term in Eq.~(\ref{eq:properaccfourth}), which cancel the effect of the first term for large distances from the beamline. In particular, we mentioned above that the first term in  Eq.~(\ref{blume}) induces an attraction with the same absolute value as the acceleration in Eq.~(\ref{eq:deflfourth}). Accordingly, if we had considered the first term in  Eq.~(\ref{blume}) only, we would have obtained an expression that would coincide with that for the geodesic deviation induced by the boosted rod given in Eq.~(\ref{eq:geodevrod}) up to a factor $2$. Therefore, we can conclude that the additional effects due to the divergence of the light beam cancel the attraction due to the reduced propagation speed of the light in the beam.

\section{Conclusion}
\label{sec:conclusions}

We analyzed the gravitational field of a focused laser beam, describing the laser beam as a solution to Maxwell's equations. We calculated the five leading orders of the metric perturbation expanded in the divergence angle $\theta$ of the beam explicitly and discussed the difference to the solutions when the laser beam is treated in the paraxial approximation. Already in the paraxial approximation, the gravitational field of a laser beam turns out to be too small to be detected with current technology \cite{raetzel_2016}. This is also the case for the effects we describe. However, they are of conceptual interest as they reveal the gravitational properties of light, and with the progress of technology, they may possibly be measurable in future experiments.

For small values of the beam waist and for $\theta=0$, which corresponds to the paraxial approximation in our case, our solution for the laser beam corresponds to the solution for the infinitely thin beam \cite{tolman_1931}. If in addition we consider the laser beam to be infinitely long, we recover the solution for an infinitely long cylinder \cite{bonnor_1969}. 

In first order in the divergence angle, we found frame dragging due to spin angular momentum of the circular polarized laser beam. This is similar to the result of \cite{strohaber_2013} for beams with orbital angular momentum. In contrast to frame dragging induced by orbital angular momentum, the effect we find decays exponentially with the distance squared from the beamline divided by the beam waist parameter $w_0$. This property coincides with the decay of the energy density of the beam. Hence, frame dragging due to the spin angular momentum of the beam is proportional to the local energy density of the beam. During the peer reviewing process for the publication of this article, the article \cite{strohaber_2018} by Strohaber appeared on the Arxiv preprint server. In the article, frame dragging due to intrinsic angular momentum including spin of light beams is derived and discussed.

The statement of \cite{tolman_1931} by Tolman et al. that a non-divergent light beam does not deflect gravitationally a co-directed parallel light beam has been recovered in different contexts: two co-directed parallel cylindrical light beams of finite radius \cite{bonnor_1969,nackoney_1973,banerjee_1975}, spinning non-divergent light beams \cite{Vsevolodovich:1989the}, non-divergent light beams in the framework of gravito-electrodynamics \cite{Faraoni:1999grav}, parallel co-propagating light-like test particles in the gravitational field of a one-dimensional light pulse \cite{raetzel_2016}. In fourth order in the divergence angle, we found a deflection of parallel co-propagating test beams. This shows that the result of \cite{tolman_1931} and \cite{bonnor_1969} only holds up to the third order in the divergence angle. This could have been expected from the fact that the group velocity of light in a Gaussian beam along the beamline is not the speed of light \cite{federov_2017,giovannini_2015}. However, the deflection of parallel co-propagating light beams by light in a focused source laser beam decays like the distribution of energy of the source beam with the distance from the beamline. This means that the effect does not persist outside of the distribution of energy given by the source laser beam like the frame dragging effect due to spin angular momentum. This is in contrast to the deflection that one obtains from a rod of matter boosted to a speed close to the speed of light. Therefore, we conclude that focused light does not simply behave like massive matter moving with the reduced velocity identified in \cite{strohaber_2013,raetzel_2016}. This is due to the divergence of the laser beam along the beamline which leads to additional contributions to the metric perturbations which do not appear in the case of the boosted rod. These additional contributions cancel the effect induced by the reduced propagation speed of light in the focused beam.

\section{Outlook}
\label{sec:outlook}

As an extension of the research presented in this article, it would be interesting to study the gravitational interaction of two parallel co-propagating focused laser beams in the description presented here. The result could be compared to the corresponding results presented in \cite{bonnor_1969,nackoney_1973,banerjee_1975}. In particular, it would be interesting to see if there exists a contribution to the gravitational interaction of the two beams that does not decay exponentially with the square of the distance between the beamlines of the beams.

It would be interesting to know if the solutions to Maxwell's equations developed in this article can be used as a basis for a quantum field theoretical description of the gravitational interaction of two laser beams in the framework of perturbative quantum gravity (PQG). Then, the effect of localization on light-light interactions could be considered for light with quantum properties. For example, in \cite{ratzel2016effect,ratzel2017effect} it is shown that the differential cross section for gravitational photon scattering can be amplified or suppressed when the scattering photons are in specific polarization entangled states initially. It would be interesting to see how this effect depends on the distance between the beams. Furthermore, in \cite{Bruschi2016}, the effect of entanglement in the position of a source of a gravitational field was investigated in the framework of semiclassical gravity. Similar questions could be considered in the framework of PQG using focused laser beams in spatial superposition states or with squeezed light as sources.

Another step from the results presented in this article into a different direction could be the consideration of a pulse of light in a focused laser mode. The framework used in this article would need to be extended to envelope functions that depend on time and the position along the beamline. Approaches for the description of such beams are given for example in \cite{ziolkowski1992propagation,salamin2002electron,april2010ultrashort,li2016fields}. An expression for the gravitational field of a focused laser pulse could be used to have a closer look at the implications of focusing for possible experiments trying to detect the gravitational field of light. In particular, the pulsed beams would produce a pulsed gravitational signal that could be detected with resonator systems like small scale gravitational wave detectors (for example \cite{Sabin:2014bua,Singh:2016xwa,Goryachev:2014yra}) or quantum optomechanical systems. 

The gravitational field of a focused laser pulse could be used as well to check the results of \cite{raetzel_2016} where the laser pulse is modeled as a one-dimensional rod of light with an energy density that is modulated as that of a plane wave. In particular, for the model used in \cite{raetzel_2016}, all gravitational effects are induced by the emission and the absorption of the light pulse alone; there is no gravitational effect related to the propagation of the pulse. This situation may change once divergence of the beam is taken into account. 

It could be worthwhile to see whether a similar solution for the gravitational field of a focused laser beam as we derived in this article could be derived considering the full coupled set of the Einstein-Maxwell equations. The resulting metric could be compared to the one in \cite{lynden2017komar} and it could be investigated if the results of \cite{lynden2017komar} about the effective gravitating mass and angular momentum can be reproduced when divergence of the beam is taken into account. It would also be interesting to consider the gravitational field of the electromagnetic field distribution used in this article to model a focused laser beam in dynamical spacetime theories with spacetime torsion like Einstein-Cartan-Theory and the Poincar\'e-Gauge-Theory of gravity \cite{Hehl1976}. In particular, we found that frame dragging due to the spin angular momentum of light is proportional to the local energy density of the beam. This is similar to the effect of spin angular momentum on test particles or fields via spacetime torsion as torsion is not a propagating degree of freedom in Einstein-Cartan-Theory and Poincar\'e-Gauge-Theory.

\section*{Acknowledgements}

We would like to thank Robert Beig, Piotr T. Chru\'sciel, Peter C. Aichelburg, David M. Fajman, Julien Fra\"isse, Lars Andersson, Jir\'i Bi\v{c}\'ak, Marius Oancea, Ralf Menzel and Martin Wilkens for interesting discussions and useful remarks. DR thanks the Humboldt foundation for its support.

\appendix
\section{Vector potential of a circularly polarized laser  beam}
\label{sec:fexpansion}

From the expansion of the field strength $F^\lambda_{\bar\alpha\bar\beta}
	= \sum_{n=0}^\infty \theta^{n} \frac{w_0 E_0}{\sqrt{2}} f^{\lambda(n)}_{\bar\alpha\bar\beta}(\xi,\chi,\theta\zeta)e^{i\frac{2}{\theta}(\zeta-\tau)}$, where $E_0=\sqrt{2}\mathcal{A}/w_0\theta$, and
the Lorenz gauge condition
\begin{equation}
	 A_{\tau} = \frac{i\theta}{2} \partial_{\tau} A_{\tau} =  \frac{i\theta}{2} \left(\partial_\xi A_\xi + \partial_{\chi} A_{\chi} + \theta\partial_{\theta\zeta} A_{\zeta} \right)  \,,
\end{equation}
we obtain a direct relation between $v_{\bar\alpha}^{\lambda(n)}$ and $f^{\lambda(n)}_{\bar\alpha\bar\beta}$ (where $\lambda$ refers to the polarization state) as 
\begin{eqnarray}\label{eq:fexprelation}
	\nonumber f^{\lambda(n)}_{\tau\zeta} &=&  \partial_\xi v^{\lambda(n-1)}_\xi + \partial_\chi v^{\lambda(n-1)}_\chi + 2\partial_{\theta\zeta} v^{\lambda(n-2)}_{\zeta} \\
	&& - \frac{i}{2} \partial_{\theta\zeta}\left(\partial_\xi v^{\lambda(n-3)}_\xi + \partial_\chi v^{\lambda(n-3)}_\chi + \partial_{\theta\zeta} v^{\lambda(n-4)}_{\zeta} \right)\;, \\
	f^{\lambda(n)}_{\tau\bar j} &=& - 2i v^{\lambda(n)}_{\bar j} + \partial_{\bar j} v^{(n-1)}_\zeta - \frac{i}{2} \partial_{\bar j}\left(\partial_\xi v^{\lambda(n-2)}_\xi + \partial_\chi v^{\lambda(n-2)}_\chi + \partial_{\theta\zeta} v^{\lambda(n-3)}_{\zeta} \right)\;,\\
	f^{\lambda(n)}_{\bar j \zeta} &=& - 2i v^{\lambda(n)}_{\bar j} + \partial_{\bar j} v^{\lambda(n-1)}_{\zeta} - \partial_{\theta\zeta} v^{\lambda(n-2)}_{\bar j} \;, \\
	f^{\lambda(n)}_{\xi\chi} &=& \partial_{\xi} v^{\lambda(n-1)}_{\chi} - \partial_{\chi} v^{\lambda(n-1)}_{\xi}\;.
\end{eqnarray}
Since the vector potential fulfills the wave equation (\ref{m}), we have that $\Box F_{\bar\alpha\bar\beta} = 0$. In particular,
\begin{eqnarray}\label{eq:parhelmf}
	\left(\partial_\xi^2+\partial_\chi^2+ 4i\partial_{\theta\zeta}\right)f^{(0)}_{\bar\alpha\bar\beta}(\xi,\chi,\theta\zeta)
	&=& 0\;,\\
	\label{eq:parhelm1f}
	\left(\partial_\xi^2+\partial_\chi^2+ 4i\partial_{\theta\zeta}\right)f^{(1)}_{\bar\alpha\bar\beta}(\xi,\chi,\theta\zeta)
	&=& 0\;,\\
	\label{eq:parhelmhigherf}\left(\partial_\xi^2+\partial_\chi^2+ 4i\partial_{\theta\zeta}\right)f^{(n)}_{\bar\alpha\bar\beta}(\xi,\chi,\theta\zeta)
	&=&	-\partial_{\theta\zeta}^2 f^{(n-2)}_{\bar\alpha\bar\beta}(\xi,\chi,\theta\zeta)\,,\;\rm{for}\,\,n> 1\,.
\end{eqnarray}
The components of the Hodge dual of the field strength tensor are given as
\begin{eqnarray}\label{eq:fexprelationdual}
	\star f^{\lambda(n)}_{\tau\zeta} = - f^{\lambda(n)}_{\xi\chi}\quad\,,\quad
	\star f^{\lambda(n)}_{\tau\xi} = -f^{\lambda(n)}_{\chi\zeta} \quad\,,\quad
	\star f^{\lambda(n)}_{\tau\chi} = f^{\lambda(n)}_{\xi\zeta}\;, \\
	\star f^{\lambda(n)}_{\xi\zeta} = -f^{\lambda(n)}_{\tau\chi} \quad\,,\quad
	\star f^{\lambda(n)}_{\chi\zeta} = f^{\lambda(n)}_{\tau\xi} \quad\,\rm{and}\quad
	\star f^{\lambda(n)}_{\xi\chi} = f^{\lambda(n)}_{\tau\zeta}\;,
\end{eqnarray}
and we obtain that a helicity eigenstate has to fulfill the conditions
\begin{eqnarray}
	\nonumber 0 &=& f^{\lambda(n)}_{\tau\zeta} + i\lambda \star f^{\lambda(n)}_{\tau\zeta} = f^{\lambda(n)}_{\tau\zeta} - i\lambda f^{\lambda(n)}_{\xi\chi}\\
	\A &=&    - i\lambda\left(\partial_{\xi} v^{\lambda(n-1)}_{\chi} - \partial_{\chi} v^{\lambda(n-1)}_{\xi}\right) + \partial_\xi  v^{\lambda(n-1)}_\xi + \partial_\chi v^{\lambda(n-1)}_\chi  + 2\partial_{\theta\zeta} v^{\lambda(n-2)}_{\zeta} \\
	 && - \frac{i}{2} \partial_{\theta\zeta}\left(\partial_\xi v^{\lambda(n-3)}_\xi + \partial_\chi v^{\lambda(n-3)}_\chi + \partial_{\theta\zeta} v^{\lambda(n-4)}_{\zeta} \right)  \;,\\
	\nonumber 0 &=& f^{\lambda(n)}_{\tau\xi} + i\lambda \star f^{\lambda(n)}_{\tau\xi} =    f^{\lambda(n)}_{\tau\xi} - i\lambda f^{\lambda(n)}_{\chi\zeta} \\
	\A &=& - 2i v^{\lambda(n)}_{\xi} + \partial_{\xi} v^{(n-1)}_\zeta - \frac{i}{2} \partial_{\xi}\left(\partial_\xi v^{\lambda(n-2)}_\xi + \partial_{\chi} v^{\lambda(n-2)}_{\chi} + \partial_{\theta\zeta} v^{\lambda(n-3)}_{\zeta} \right)\\
	&& -i\lambda\left(- 2i v^{\lambda(n)}_{\chi} + \partial_{\chi} v^{\lambda(n-1)}_{\zeta} - \partial_{\theta\zeta} v^{\lambda(n-2)}_{\chi}\right)\;,\\
	\nonumber 0 &=& f^{\lambda(n)}_{\tau\chi} + i\lambda \star f^{\lambda(n)}_{\tau\chi} =    f^{\lambda(n)}_{\tau\chi} + i\lambda f^{\lambda(n)}_{\xi\zeta} \\
	\A &=& - 2i v^{\lambda(n)}_{\chi} + \partial_{\chi} v^{(n-1)}_\zeta - \frac{i}{2} \partial_{\chi}\left(\partial_\xi v^{\lambda(n-2)}_\xi + \partial_{\chi} v^{\lambda(n-2)}_{\chi} + \partial_{\theta\zeta} v^{\lambda(n-3)}_{\zeta} \right)\\
	&& + i\lambda\left(- 2i v^{\lambda(n)}_{\xi} + \partial_{\xi} v^{\lambda(n-1)}_{\zeta} - \partial_{\theta\zeta} v^{\lambda(n-2)}_{\xi}\right)\;,\\
	0 &=& f^{\lambda(n)}_{\xi \zeta}   + i\lambda \star  f^{\lambda(n)}_{\xi \zeta} = f^{\lambda(n)}_{\xi \zeta}   - i\lambda f^{\lambda(n)}_{\tau\chi} = -i\lambda \left(  f^{\lambda(n)}_{\tau\chi} + i\lambda f^{\lambda(n)}_{\xi \zeta} \right)\;, \\
	0 &=& f^{\lambda(n)}_{\chi \zeta}   + i\lambda \star  f^{\lambda(n)}_{\chi\zeta} = f^{\lambda(n)}_{\xi \zeta}   + i\lambda f^{\lambda(n)}_{\tau\xi} = i\lambda \left(  f^{\lambda(n)}_{\tau\xi} - i\lambda f^{\lambda(n)}_{\chi \zeta} \right)\;, \\
	0 &=& f^{\lambda(n)}_{\xi\chi} + i\lambda \star f^{\lambda(n)}_{\xi\chi} = f^{\lambda(n)}_{\xi\chi} + i\lambda f^{\lambda(n)}_{\tau\zeta} = i\lambda\left(f^{\lambda(n)}_{\tau\zeta} - i\lambda f^{\lambda(n)}_{\xi\chi} \right) \,,
\end{eqnarray}
where the last three conditions are fulfilled if the first three conditions are fulfilled. The remaining conditions can be rewritten as
\begin{eqnarray}\label{eq:fexpcond1}
	\nonumber 0 &=&  \left(\partial_\xi + i\lambda\partial_\chi\right)  \left( v^{\lambda(n-1)}_\xi  - i\lambda v^{\lambda(n-1)}_{\chi}\right)    + 2\partial_{\theta\zeta} v^{\lambda(n-2)}_{\zeta} \\
	 && - \frac{i}{2} \partial_{\theta\zeta}\left(\partial_\xi v^{\lambda(n-3)}_\xi + \partial_\chi v^{\lambda(n-3)}_\chi + \partial_{\theta\zeta} v^{\lambda(n-4)}_{\zeta} \right)\;,  \\
	\nonumber 0  &=& - 2i\left( v^{\lambda(n)}_{\xi} - i\lambda v^{\lambda(n)}_{\chi} \right) + \left(\partial_{\xi} - i\lambda\partial_\chi \right) v^{(n-1)}_\zeta \\\label{second}
	&& - \frac{i}{2} \partial_{\xi}\left(\partial_\xi v^{\lambda(n-2)}_\xi + \partial_{\chi} v^{\lambda(n-2)}_{\chi} \right) + i\lambda  \partial_{\theta\zeta} v^{\lambda(n-2)}_{\chi} - \frac{i}{2} \partial_{\xi}\partial_{\theta\zeta} v^{\lambda(n-3)}_{\zeta} \;,  \\
	\nonumber 0  &=& - 2i \left(v^{\lambda(n)}_{\xi} - i\lambda v^{\lambda(n)}_{\chi}\right)  + \left( \partial_\xi - i\lambda\partial_{\chi} \right) v^{(n-1)}_\zeta \\\label{third}
	&& - \frac{\lambda}{2} \partial_{\chi}\left(\partial_\xi v^{\lambda(n-2)}_\xi + \partial_{\chi} v^{\lambda(n-2)}_{\chi} \right)  - \partial_{\theta\zeta} v^{\lambda(n-2)}_{\xi}  - \frac{\lambda}{2} \partial_{\chi}\partial_{\theta\zeta} v^{\lambda(n-3)}_{\zeta} \;.
\end{eqnarray}
The sum and the difference of Eq.~\ref{second} and Eq.~\ref{third} lead to
\begin{eqnarray}
	\nonumber 0  &=& - 4i\left( v^{\lambda(n)}_{\xi} - i\lambda v^{\lambda(n)}_{\chi} \right) + 2\left(\partial_{\xi} - i\lambda\partial_\chi \right) v^{(n-1)}_\zeta - \frac{i}{2}\left(\partial_{\xi} - i\lambda \partial_\chi\right)\left(\partial_\xi v^{\lambda(n-2)}_\xi + \partial_{\chi} v^{\lambda(n-2)}_{\chi} \right) \\
	\label{eq:fexpcond2} &&  - \partial_{\theta\zeta} \left(v^{\lambda(n-2)}_{\xi} - i\lambda   v^{\lambda(n-2)}_{\chi}\right) - \frac{i}{2}\left( \partial_{\xi}  - i\lambda\partial_\chi \right) \partial_{\theta\zeta} v^{\lambda(n-3)}_{\zeta}   \;,
\end{eqnarray}
and
\begin{eqnarray}
		\nonumber 0  &=& - \frac{i}{2}\left(\partial_\xi + i\lambda \partial_{\chi}\right)\left(\partial_\xi v^{\lambda(n-2)}_\xi + \partial_{\chi} v^{\lambda(n-2)}_{\chi} \right)  + \partial_{\theta\zeta} \left(v^{\lambda(n-2)}_{\xi} + i\lambda v^{\lambda(n-2)}_\chi\right)\\
	\A &&   - \frac{i}{2}\left(\partial_\xi + i\lambda \partial_{\chi}\right) \partial_{\theta\zeta} v^{\lambda(n-3)}_{\zeta} \\
	\A &=& - \frac{i}{2}\left(\partial_\xi + i\lambda \partial_{\chi}\right)\left(\partial_\xi v^{\lambda(n-2)}_\xi + \partial_{\chi} v^{\lambda(n-2)}_{\chi} \right)  + \frac{i}{4} \left(\partial_\xi^2+\partial_\chi^2\right) \left(v^{\lambda(n-2)}_{\xi}  +    i\lambda   v^{\lambda(n-2)}_{\chi}\right) \\
	\A && + \frac{i}{4}\partial_{\theta\zeta}^2\left(v^{\lambda(n-4)}_{\xi} + i\lambda   v^{\lambda(n-4)}_{\chi}\right)   - \frac{i}{2}\left(\partial_\xi + i\lambda \partial_{\chi}\right) \partial_{\theta\zeta} v^{\lambda(n-3)}_{\zeta} \\
	\A &=& - \frac{i}{4}\left(\partial_\xi + i\lambda \partial_{\chi}\right)^2\left( v^{\lambda(n-2)}_\xi - i\lambda v^{\lambda(n-2)}_{\chi} \right)   \\
	\label{eq:fexpcond3} && - \frac{i}{2}\left(\partial_\xi + i\lambda \partial_{\chi}\right) \partial_{\theta\zeta} v^{\lambda(n-3)}_{\zeta}  + \frac{i}{4}\partial_{\theta\zeta}^2\left(v^{\lambda(n-4)}_{\xi} + i\lambda   v^{\lambda(n-4)}_{\chi}\right)   \,,
\end{eqnarray}
respectively. For the leading/zeroth order envelope function, we find from Eq.~(\ref{eq:fexpcond2}) that $v^{\lambda(0)}_{\xi} = i\lambda v^{\lambda(0)}_{\chi}$. For the first order envelope function, we obtain from Eq.~(\ref{eq:fexpcond1}) the condition
\begin{eqnarray}
	\A 0 &=&  \left(\partial_\xi + i\lambda\partial_\chi\right)  \left( v^{\lambda(0)}_\xi  - i\lambda v^{\lambda(0)}_{\chi}\right) \;,
\end{eqnarray}
which is fulfilled for $v^{\lambda(0)}_{\xi} = i\lambda v^{\lambda(0)}_{\chi}$. Furthermore from Eq.~(\ref{eq:fexpcond2}), we find the condition
\begin{eqnarray}\label{eq:condfirstorder}
 0  &=& - 2i\left( v^{\lambda(1)}_{\xi} - i\lambda v^{\lambda(1)}_{\chi} \right) + \left(\partial_{\xi} - i\lambda\partial_\chi \right) v^{(0)}_\zeta \,.
\end{eqnarray}
For the second order, we obtain from Eq.~(\ref{eq:fexpcond1})
\begin{eqnarray}
	\A 0 &=&  \left(\partial_{\xi} +  i\lambda \partial_\chi\right) \left( v^{\lambda(1)}_\xi - i  \lambda v^{\lambda(1)}_{\chi} \right) + 2\partial_{\theta\zeta} v^{\lambda(0)}_{\zeta} \\
	\A &=& - \frac{i}{2}\Delta_\mathrm{2d} v^{\lambda(0)}_{\zeta} + 2\partial_{\theta\zeta} v^{\lambda(0)}_{\zeta}\;,
\end{eqnarray}
which is always fulfilled since $v^{\lambda(0)}_{\zeta}$ fulfills Eq.~(\ref{eq:parhelm}). Additionally from Eq.~(\ref{eq:fexpcond2}) and with $v^{\lambda(0)}_{\xi} = i\lambda v^{\lambda(0)}_{\chi}$, we find the condition
\begin{eqnarray}
	\A 0  &=& - 2i\left( v^{\lambda(2)}_{\xi} - i\lambda v^{\lambda(2)}_{\chi} \right) + \left(\partial_{\xi} - i\lambda\partial_\chi \right) v^{(1)}_\zeta - \frac{i}{4}\left(\partial_{\xi} - i\lambda \partial_\chi\right)\left(i\lambda\partial_\xi  + \partial_{\chi}  \right)v^{\lambda(0)}_{\chi} \\
	&=&  - 2i \left( v^{\lambda(2)}_{\xi}  -i\lambda v^{\lambda(2)}_{\chi}\right) + \left(\partial_{\xi}-i\lambda \partial_{\chi}\right) v^{(1)}_\zeta + \frac{\lambda}{4}\left(\partial_{\xi}-i\lambda \partial_{\chi}\right)^2 v^{\lambda(0)}_{\chi}\,.
\end{eqnarray}
Assuming $v^{\lambda(2)}_{\xi} = i\lambda v^{\lambda(2)}_{\chi}$, we find that the first term in the condition vanishes and we can solve for $v^{(1)}_\zeta$ as
\begin{eqnarray}\label{eq:condsecond}
	 v^{(1)}_\zeta	&=&    - \frac{\lambda}{4}\left(\partial_{\xi}-i\lambda \partial_{\chi}\right) v^{\lambda(0)}_{\chi}\,.
\end{eqnarray}
The condition in Eq.~(\ref{eq:fexpcond3}) is automatically fulfilled in second order due to $v^{\lambda(2)}_{\xi} = i\lambda v^{\lambda(2)}_{\chi}$. For the third order, we find from Eq.~(\ref{eq:fexpcond1})
\begin{eqnarray}
	 0 &=&   \partial_{\theta\zeta} v^{\lambda(1)}_{\zeta} + \frac{\lambda}{4} \partial_{\theta\zeta}\left(\partial_\xi - i\lambda \partial_\chi \right)v^{\lambda(0)}_\chi \;,
\end{eqnarray}
which is just the $\theta\zeta$-derivative of Eq.~(\ref{eq:condsecond}). From Eq.~(\ref{eq:fexpcond2}) follows that
\begin{eqnarray}
	\A 0  &=& - 4i\left( v^{\lambda(3)}_{\xi} - i\lambda v^{\lambda(3)}_{\chi} \right) + 2\left(\partial_{\xi} - i\lambda\partial_\chi \right) v^{(2)}_\zeta - \frac{i}{2}\left(\partial_{\xi} - i\lambda \partial_\chi\right)\left(\partial_\xi v^{\lambda(1)}_\xi + \partial_{\chi} v^{\lambda(1)}_{\chi} \right) \\
	\A &&  - \partial_{\theta\zeta} \left(v^{\lambda(1)}_{\xi} - i\lambda   v^{\lambda(1)}_{\chi}\right) - \frac{i}{2}\left( \partial_{\xi}  - i\lambda\partial_\chi \right) \partial_{\theta\zeta} v^{\lambda(0)}_{\zeta}\;,\\
	&=& - 4i\left( v^{\lambda(3)}_{\xi} - i\lambda v^{\lambda(3)}_{\chi} \right) + 2\left(\partial_{\xi} - i\lambda\partial_\chi \right) v^{(2)}_\zeta - \frac{i}{2}\left(\partial_{\xi} - i\lambda \partial_\chi\right)\left(\partial_\xi v^{\lambda(1)}_\xi + \partial_{\chi} v^{\lambda(1)}_{\chi} \right)\;,
\end{eqnarray}
where we used Eq.~(\ref{eq:condfirstorder}). The last condition of third order comes from Eq.~(\ref{eq:fexpcond3}) as
\begin{eqnarray}
	\A 0 &=& - \frac{i}{4}\left(\partial_\xi + i\lambda \partial_{\chi}\right)^2\left( v^{\lambda(1)}_\xi - i\lambda v^{\lambda(1)}_{\chi} \right)  - \frac{i}{2}\left(\partial_\xi + i\lambda \partial_{\chi}\right) \partial_{\theta\zeta} v^{\lambda(0)}_{\zeta}\\
	  &=& - \frac{1}{8}\left(\partial_\xi + i\lambda \partial_{\chi}\right)\Delta_\mathrm{2d} v^{\lambda(0)}_{\zeta}  - \frac{i}{2}\left(\partial_\xi + i\lambda \partial_{\chi}\right) \partial_{\theta\zeta} v^{\lambda(0)}_{\zeta} \;,   
\end{eqnarray}
which is fulfilled since Eq.~(\ref{eq:parhelmschroe}) has to hold. In fourth order, we find from 
(\ref{eq:fexpcond1})
\begin{eqnarray}
	\A  0 &=&  \left(\partial_\xi + i\lambda\partial_\chi\right)  \left( v^{\lambda(3)}_\xi  - i\lambda v^{\lambda(3)}_{\chi}\right)    + 2\partial_{\theta\zeta} v^{\lambda(2)}_{\zeta} \\
	\A  && - \frac{i}{2} \partial_{\theta\zeta}\left(\partial_\xi v^{\lambda(1)}_\xi + \partial_\chi v^{\lambda(1)}_\chi + \partial_{\theta\zeta} v^{\lambda(0)}_{\zeta} \right)\\
	\A  &=&  -\Delta_\mathrm{2d}  \left( \frac{i}{2} v^{(2)}_\zeta  + \frac{1}{8} \left(\partial_\xi v^{\lambda(1)}_\xi + \partial_{\chi} v^{\lambda(1)}_{\chi} \right)\right) + 2\partial_{\theta\zeta} v^{\lambda(2)}_{\zeta} \\
	\A  && - \frac{i}{2} \partial_{\theta\zeta}\left(\partial_\xi v^{\lambda(1)}_\xi + \partial_\chi v^{\lambda(1)}_\chi + \partial_{\theta\zeta} v^{\lambda(0)}_{\zeta} \right) \\
	\A  &=&  -\frac{i}{2}\left(\Delta_\mathrm{2d} v^{\lambda(2)}_\zeta + 4i\partial_{\theta\zeta} v^{\lambda(2)}_{\zeta} + \partial_{\theta\zeta}^2 v^{\lambda(0)}_{\zeta}\right)\\
	 && -\frac{1}{8} \left( \Delta_\mathrm{2d} \left( \partial_\xi v^{\lambda(1)}_\xi + \partial_{\chi} v^{\lambda(1)}_{\chi} \right)  + 4i  \partial_{\theta\zeta}\left( \partial_\xi v^{\lambda(1)}_\xi + \partial_\chi v^{\lambda(1)}_\chi \right)\right)\;,
\end{eqnarray}
which is satisfied due to Eqs.~(\ref{eq:parhelm1}) and (\ref{eq:parhelmhigher}). From Eq.~(\ref{eq:fexpcond2}), we obtain in fourth order
\begin{eqnarray}
	\nonumber 0  &=& - 4i\left( v^{\lambda(4)}_{\xi} - i\lambda v^{\lambda(4)}_{\chi} \right) + 2\left(\partial_{\xi} - i\lambda\partial_\chi \right) v^{(3)}_\zeta - \frac{i}{2}\left(\partial_{\xi} - i\lambda \partial_\chi\right)\left(\partial_\xi v^{\lambda(2)}_\xi + \partial_{\chi} v^{\lambda(2)}_{\chi} \right) \\
	 &&  - \partial_{\theta\zeta} \left(v^{\lambda(2)}_{\xi} - i\lambda   v^{\lambda(2)}_{\chi}\right) - \frac{i}{2}\left( \partial_{\xi}  - i\lambda\partial_\chi \right) \partial_{\theta\zeta} v^{\lambda(1)}_{\zeta}    \;.
\end{eqnarray}
Assuming $v^{\lambda(4)}_{\xi} = i\lambda v^{\lambda(4)}_{\chi}$ and taking into account $v^{\lambda(2)}_{\xi} = i\lambda v^{\lambda(2)}_{\chi}$, which we assumed before, we obtain
\begin{eqnarray}
	 0  &=&  2\left(\partial_{\xi} - i\lambda\partial_\chi \right) v^{(3)}_\zeta - \frac{i}{2}\left(\partial_{\xi} - i\lambda \partial_\chi\right)\left(\partial_\xi v^{\lambda(2)}_\xi + \partial_{\chi} v^{\lambda(2)}_{\chi} \right)   - \frac{i}{2}\left( \partial_{\xi}  - i\lambda\partial_\chi \right) \partial_{\theta\zeta} v^{\lambda(1)}_{\zeta}  \;.  
\end{eqnarray}
With Eq.~(\ref{eq:condsecond}), we obtain that
\begin{eqnarray}\label{eq:condfourth}
	   v^{(3)}_\zeta  &=&  -\frac{\lambda}{4}\left( \partial_\xi -i\lambda \partial_{\chi}  \right)\left( v^{\lambda(2)}_{\chi} + \frac{i}{4} \partial_{\theta\zeta}  v^{\lambda(0)}_{\chi} \right)\;.
\end{eqnarray}
Again with Eq.~(\ref{eq:condsecond}), we can check that the higher order Helmholtz equation (\ref{eq:parhelmhigher}) is fulfilled by $v^{(3)}_\zeta$ given in (\ref{eq:condfourth}). The last condition that we have to check is the fourth order condition in Eq.~(\ref{eq:fexpcond3}), which becomes
\begin{eqnarray}
	\A 0 &=& - \frac{i}{4}\left(\partial_\xi + i\lambda \partial_{\chi}\right)^2\left( v^{\lambda(2)}_\xi - i\lambda v^{\lambda(2)}_{\chi} \right)   \\
	 && - \frac{i}{2}\left(\partial_\xi + i\lambda \partial_{\chi}\right) \partial_{\theta\zeta} v^{\lambda(1)}_{\zeta}  + \frac{i}{4}\partial_{\theta\zeta}^2\left(v^{\lambda(0)}_{\xi} + i\lambda   v^{\lambda(0)}_{\chi}\right) \,,
\end{eqnarray}
which may be written as, using $v^{\lambda(2)}_{\xi} = i\lambda v^{\lambda(2)}_{\chi}$ and $v^{\lambda(0)}_{\xi} = i\lambda v^{\lambda(0)}_{\chi}$\;,
\begin{eqnarray}
	 0 &=& - \frac{i}{2}\left(\partial_\xi + i\lambda \partial_{\chi}\right) \partial_{\theta\zeta} v^{\lambda(1)}_{\zeta}  - \frac{\lambda}{2}\partial_{\theta\zeta}^2 v^{\lambda(0)}_{\chi} \;,
\end{eqnarray}
and is fulfilled due to Eqs.~(\ref{eq:condsecond}) and (\ref{eq:parhelm}). We conclude that a 
vector potential for a circularly polarized laser  beam up to fourth order in the divergence angle $\theta$ is given by Eqs.~(\ref{eq:parhelm}), (\ref{eq:parhelm1}) and (\ref{eq:parhelmhigher}), Eqs.~(\ref{eq:condsecond}) and (\ref{eq:condfourth}) and the additional sufficient conditions $v^{\lambda(2n)}_{\xi} = i\lambda v^{\lambda(2n)}_{\chi}$ and $v^{\lambda(2n)}_{\zeta}=0$ for $n=0,1,2$ and $v^{\lambda(2n+1)}_{\xi} = 0 = v^{\lambda(2n+1)}_{\chi}$ for $n=0,1$. 

Starting from $v^{(0)}_{\bar\alpha} = \epsilon_{\bar\alpha} v_0$, where $\epsilon_{\bar\alpha}=w_0(0,1,-\lambda i,0)/\sqrt{2}$ and the solutions of even orders that can be found in  \cite{Salamin:2007fie},
\begin{eqnarray}
	v^{(0)}_{\bar\alpha}(\xi,\chi,\theta\zeta)
	&=& \epsilon^{(0)}_{\bar\alpha}v_0(\xi,\chi,\theta\zeta)\;,\\
	v^{(2)}_{\bar\alpha}(\xi,\chi,\theta\zeta)
	&=&	\frac{\mu(\theta\zeta)}{2}\left(1 - \frac{1}{2}\mu(\theta\zeta)^2\rho^4 \right)v^{(0)}_{\bar\alpha}(\xi,\chi,\theta\zeta)\;,\\
	v^{(4)}_{\bar\alpha}(\xi,\chi,\theta\zeta)
	&=&	\frac{\mu(\theta\zeta)^2}{16}  \left(6  - 3\mu(\theta\zeta)^2\rho^4 - 2\mu(\theta\zeta)^3\rho^6 + \frac{1}{2}\mu(\theta\zeta)^4 \rho^8 \right)v^{(0)}_{\bar\alpha}(\xi,\chi,\theta\zeta) \;,
\end{eqnarray}
where $v_0(\xi,\chi,\theta\zeta) = \mu(\theta\zeta)e^{-\mu(\theta\zeta)\rho^2}$. This leads to the expressions for the odd orders
\begin{eqnarray}
	v^{(1)}_{\zeta}(\xi,\chi,\theta\zeta)
	&=& -\frac{iw_0\mu(\theta\zeta)}{2\sqrt{2}}\left(\xi - i\lambda\chi\right)v_0(\xi,\chi,\theta\zeta)\;,\\
	v^{(3)}_{\zeta}(\xi,\chi,\theta\zeta)
	&=&	\frac{\mu(\theta\zeta)}{4}\left(4 +  \mu(\theta\zeta)\rho^2 - \mu(\theta\zeta)^2\rho^4 \right)v_\zeta^{(1)}(\xi,\chi,\theta\zeta\;.
\end{eqnarray}
\vfill

\section{Poynting vector, Maxwell stress tensor and energy}
\label{sec:pointing}

For the vector potential of a circularly polarized laser  beam given by Eq.~(\ref{eq:circpolenv}), the energy density, the Poynting vector and the stress tensor components are given as
\small{\begin{eqnarray}
	\nonumber \mathcal{E}^\lambda 
	&=& \mathcal{E}^{(0)} \left[ 
			1 + \frac{|\mu|^2\theta^2}{2}\bigg(1 + |\mu|^2\Big(2 - (4|\mu|^2 - 3)\rho^2\Big)\rho^2\bigg)  + \frac{|\mu|^2\theta^4}{16}\bigg( -3 + 2|\mu|^2(4 - \rho^2 - \rho^4) \right.\\
	&&	\left. + 4|\mu|^4\big(4 - \rho^2 - 5\rho^4\big)\rho^2  + 2|\mu|^6\big(8+52\rho^2 + 9\rho^4\big)\rho^4 - 48|\mu|^8\big(2+\rho^2\big)\rho^6 + 32|\mu|^{10}\rho^8 \bigg) \right]\;, \\
	\A S^\lambda_\xi /c 
	&=&  \mathcal{E}^{(0)} \theta |\mu|^2\bigg[  (\theta \zeta  \xi +\lambda  \chi ) \\
	&& - \frac{\theta^2}{4}   \bigg(\lambda\chi - 2|\mu|^2\Big((2-\rho^2)\theta\zeta\xi + 2(1-\rho^2)\lambda\chi + (\theta\zeta\xi + \lambda\chi)(4 + 3\rho^2 - 4|\mu|^2 \rho^2)|\mu|^2\rho^2 \Big)\bigg)\bigg]\;,\\\A
	\A S^\lambda_\chi /c
	 &=&	 - \lambda \mathcal{E}^{(0)}  \theta |\mu|^2\bigg[  ( \xi - \theta \zeta \lambda\chi ) \\
	&& -  \frac{\theta^2}{4}   \bigg(\xi - 2|\mu|^2\Big(2(1-\rho^2)\xi - (2-\rho^2)\theta\zeta\lambda\chi + (\xi - \theta\zeta\lambda\chi)(4 + 3\rho^2 - 4|\mu|^2 \rho^2 )|\mu|^2\rho^2 \Big) \bigg)\bigg]\;,\\
	S^\lambda_\zeta /c 
	&=&	\mathcal{E}^\lambda - \frac{1}{2}\mathcal{E}^{(0)} (\theta\rho|\mu|)^2\left[ 1 
	+ \left(\frac{\theta|\mu|}{2}\right)^2\bigg(4 - 3\rho^2 + (8+6\rho^2-8\rho^2|\mu|^2)\rho^2|\mu|^2 \bigg) \right]\;, \\
	\A \sigma^\lambda_{\xi\xi}
	&=&	\mathcal{E}^{(0)} \theta^2 |\mu|^4 (\theta \zeta  \xi + \lambda  \chi )\bigg[(\theta\zeta\xi + \lambda\chi)     + \frac{\theta^2}{2}  \bigg( - \lambda\chi   + 3|\mu|^2(\theta \zeta \xi + \lambda\chi )  \\
	&& + \rho^2|\mu|^2 \left(-2\lambda\chi - 2 (1 - 3|\mu|^2)(\theta\zeta \xi + \lambda\chi) +\left(3 - 4|\mu|^2\right) \rho^2 |\mu|^2 (\theta \zeta  \xi +\lambda  \chi )\right)\bigg) \bigg] \;,\\\A
	\sigma^\lambda_{\chi\chi}
	&=&	\mathcal{E}^{(0)} \theta^2 |\mu|^4 (  \xi - \theta \zeta \lambda  \chi )\bigg[(\xi - \theta \zeta \lambda\chi)     + \frac{\theta^2}{2}  \bigg( - \xi   + 3|\mu|^2(\xi - \theta \zeta \lambda\chi )  \\
	&& + \rho^2|\mu|^2 \left(-2\xi - 2 (1 - 3|\mu|^2)(\xi - \theta \zeta\lambda\chi) +\left(3 - 4|\mu|^2\right) \rho^2 |\mu|^2 (\xi - \theta \zeta\lambda  \chi )\right)\bigg) \bigg]\;, \\\A
	\sigma^\lambda_{\xi\chi}
	&=&	 \mathcal{E}^{(0)} \lambda  \theta^2 |\mu|^4 \bigg[ (\theta \zeta  \xi +\lambda  \chi ) (\theta \zeta  \lambda  \chi -\xi ) - \frac{\theta^2}{4}  \bigg(\theta\zeta\left(6|\mu|^2 - 1\right)\left(\xi^2-\chi^2\right) + 4(3|\mu|^2-2)\lambda\xi\chi   \\
	\A && + 2 \rho^2  \bigg( 3\theta \zeta \left( 2|\mu|^2 - 1 \right) |\mu|^2 \left(\xi^2-\chi^2\right)  +  2\left(6|\mu|^4 - 6|\mu|^2 + 1\right) \lambda\xi\chi  \\
	&&  +\left(4 |\mu|^2 - 3\right) \rho^2 |\mu|^4 (\theta \zeta  \xi +\lambda  \chi ) (\theta \zeta  \lambda  \chi -\xi )\bigg)\bigg)\bigg]	\;,\\
	\sigma^\lambda_{\xi\zeta}
	&=&  S^\lambda_\xi /c  - \mathcal{E}^{(0)} \frac{\theta^3}{2}   (\theta \zeta  \xi +\lambda  \chi )|\mu|^4\rho^2 \;,\\
	\sigma^\lambda_{\chi\zeta}
	&=&  S^\lambda_\chi /c  +  \lambda \mathcal{E}^{(0)} \frac{\theta^3}{2}  (\xi -  \theta \zeta  \lambda \chi )|\mu|^4\rho^2 \;,\\
	\sigma^\lambda_{\zeta\zeta} 
	&=&  \mathcal{E}^\lambda - \mathcal{E}^{(0)} (\theta\rho|\mu|)^2\left[ 1 
	+ \left(\frac{\theta|\mu|}{2}\right)^2 \bigg(4 - 4\rho^2 + (8+6\rho^2-8\rho^2|\mu|^2)\rho^2|\mu|^2 \bigg)\right] \;,
\end{eqnarray}}
where $\mathcal{E}^{(0)} = \varepsilon_0 w_0^2 E_0^2 |v_0|^2 = 2P_0 |v_0|^2 /(\pi c) $.
\vfill

\section{The projected solution}
\label{sec:projsol}

Following the second option to construct the field strength tensor for a circularly polarized beam described in Sec.~\ref{sec:fieldstrength}, we start from the zeroth order envelope function $v^{(0)}_{\bar\alpha} = \epsilon_{\bar\alpha}^{(0)} v_0$, where $\epsilon_{\bar\alpha}^{(0)}=(0,1,-\lambda i,0)w_0/\sqrt{2}$. We define cylindrical coordinates $(\rho,\phi,\zeta)$ such that $\xi = \rho \cos\phi$ and $\chi = \rho \sin\phi$. Then, the components of the field strength tensor of the helicity eigenfunction $F^{\lambda,\mathrm{pro}}_{\bar{\alpha}\bar{\beta}}=(1+\lambda\Lambda)F^{\lambda}_{\bar{\alpha}\bar{\beta}}/2$ become
\begin{eqnarray}\label{eq:fieldstrength}
	\nonumber F^{\lambda,\mathrm{pro}}_{\tau \xi} =  -i\lambda F^{\lambda,\mathrm{pro}}_{\zeta\chi}  &=& -i w_0^2 E_0 v_0\, e^{i\frac{2}{\theta}(\zeta-\tau)}\left[1 + \left(\frac{\theta \mu \rho}{2}\right)^2 \left(2 + e^{-2i\lambda\phi} - \mu\rho^2\right)\right. \\
	&& \left. + \left(\frac{\theta \mu \rho}{2}\right)^4 \left(6 + 4 e^{-2i\lambda\phi} - \left(4 + e^{-2i\lambda\phi}\right)\mu\rho^2 + \frac{1}{2}\mu^2\rho^4 \right)\right] \;,\\
     \nonumber F^{\lambda,\mathrm{pro}}_{\tau \chi} =  -i\lambda F^{\lambda,\mathrm{pro}}_{\xi\zeta}  &=& -\lambda w_0^2 E_0 v_0\, e^{i\frac{2}{\theta}(\zeta-\tau)}\left[1 + \left(\frac{\theta \mu \rho}{2}\right)^2 \left(2 - e^{-2i\lambda\phi} - \mu\rho^2\right)\right. \\
	&& \left. + \left(\frac{\theta \mu \rho}{2}\right)^4 \left(6 - 4 e^{-2i\lambda\phi} - \left(4 - e^{-2i\lambda\phi}\right)\mu\rho^2 + \frac{1}{2}\mu^2\rho^4 \right)\right]\;, \\
     F^{\lambda,\mathrm{pro}}_{\tau \zeta} =  -i\lambda F^{\lambda,\mathrm{pro}}_{\chi\xi} &=& -   w_0^2 E_0 v_0\, e^{i\frac{2}{\theta}(\zeta-\tau)} \theta \mu\rho e^{-i\lambda\phi} \left[ 1 + \left(\frac{\theta\mu\rho}{2}\right)^2(3 - \mu\rho^2)\right]\;.
\end{eqnarray}
Since $\Lambda F_{\bar{\alpha}\bar{\beta}}=-i\star F_{\bar{\alpha}\bar{\beta}}$, the projection  $(1+\lambda\Lambda)/2$ is equivalent to adding the dual field of $F_{\bar{\alpha}\bar{\beta}}$. In the approach of complex source points presented in \cite{Cullen:1979com}, adding the dual corresponds to adding a magnetic dipole to the electric dipole that would create $F_{\bar{\alpha}\bar{\beta}}$. In contrast to \cite{Cullen:1979com}, we add the dual with a phase shift of $-\pi/2$ to add $-i\star F_{\bar{\alpha}\bar{\beta}}$ and not just $\star F_{\bar{\alpha}\bar{\beta}}$ \footnote{Note that this symmetrization of the field strength tensor done in \cite{Cullen:1979com} is also performed in \cite{Davis:1979theo} without giving reference to a magnetic dipole moment.}. 
\vfill

\subsection{Poynting vector, Maxwell stress tensor and energy}

For the field strength tensor $F^{\lambda,\mathrm{pro}}_{\bar{\alpha}\bar{\beta}}$ of a circularly polarized laser  beam given in Eq.~(\ref{eq:fieldstrength}), the energy density, the Poynting vector and the stress tensor components are given as
\small{\begin{eqnarray}
	\nonumber \mathcal{E}^\lambda 
	&=& \mathcal{E}^{(0)} \left[ 
			1 + \frac{\left(\theta\rho|\mu|\right)^2}{2}\left(2+(4|\mu|^2-3)(1-\rho^2|\mu|^2)\right) \right.\\
	&&	\left. + \frac{\left(\theta\rho|\mu|\right)^4}{16}\left(96|\mu|^4-72|\mu|^2 + 5 - 2\rho^2|\mu|^2\left(2(32|\mu|^4-36|\mu|^2+8)-(4|\mu|^2-3)^2\rho^2|\mu|^2\right)\right)
		\right]\;, \\
	 S^\lambda_\xi /c 
	&=& \mathcal{E}^{(0)} \theta |\mu|^2\left[  (\theta \zeta  \xi +\lambda  \chi ) + \frac{\left(\theta\rho|\mu|\right)^2}{2}   \left(  (6 |\mu|^2 - 2)\theta \zeta  \xi  +    (6 |\mu|^2 - 3)\lambda  \chi -  (4 |\mu|^2 -3)  (\theta \zeta  \xi +\lambda  \chi )\rho ^2|\mu|^2\right)\right]\;,\\
	S^\lambda_\chi /c 
	&=&	-\lambda \mathcal{E}^{(0)} \theta |\mu|^2\left[  (  \xi - \lambda \theta\zeta \chi ) + \frac{\left(\theta\rho|\mu|\right)^2}{2}   \left(  (6 |\mu|^2 - 3)   \xi  -    (6 |\mu|^2 - 2)\lambda \theta\zeta \chi -  (4 |\mu|^2 -3)  (  \xi - \lambda \theta \zeta \chi )\rho ^2|\mu|^2\right)\right]\;,\\
	S^\lambda_\zeta /c
	&=& E^\lambda - \frac{1}{2}\mathcal{E}^{(0)} (\theta\rho|\mu|)^2\left[ 1 
	+ \left(\frac{\theta\rho|\mu|}{2}\right)^2 2 \left(2|\mu|^2 + \frac{1}{2} + (4|\mu|^2-3)\left(1-\rho^2|\mu|^2\right)\right) 	\right] \;,\\
	\sigma^\lambda_{\xi\xi}
	&=&	\mathcal{E}^{(0)} \theta ^2 |\mu|^4 (\theta \zeta  \xi +\lambda  \chi )\bigg[(\theta \zeta  \xi +\lambda  \chi )\\
	&&  + \frac{(\theta\rho |\mu|)^2}{2}   \left(  (8 |\mu|^2-3)\theta \zeta  \xi + (8 |\mu|^2-5)\lambda  \chi   -(4 |\mu|^2-3) (\theta \zeta  \xi +\lambda  \chi )\rho^2|\mu|^2\right)\bigg] \;,\\\A
	\sigma^\lambda_{\chi\chi}
	&=&	\mathcal{E}^{(0)} \theta ^2 |\mu|^4 (  \xi  - \lambda \theta \zeta \chi )\bigg[(  \xi - \lambda \theta \zeta \chi )\\
	&&  + \frac{(\theta\rho |\mu|)^2}{2}   \left(  (8 |\mu|^2 - 5)\xi  -  (8 |\mu|^2 - 3)\lambda \theta \zeta \chi  -(4 |\mu|^2-3) (  \xi - \lambda \theta \zeta \chi )\rho^2|\mu|^2\right)\bigg] \;,\\\A
	\sigma^\lambda_{\xi\chi}
	&=&	 \mathcal{E}^{(0)} \lambda  \theta^2 |\mu|^4 \bigg[ (\theta \zeta  \xi +\lambda  \chi ) (\theta \zeta  \lambda  \chi -\xi ) -\frac{1}{2} \theta^2 \rho^2 \bigg( (\theta \zeta  \lambda  \chi -\xi ) (\theta \zeta  \xi +\lambda  \chi ) (4 |\mu|^2 -3)\rho^2|\mu|^4 \\
	&&  + 4 \theta \zeta  \xi ^2 (2 |\mu|^2-1) |\mu|^2 - 4 \theta \zeta  \chi ^2 (2 |\mu|^2-1) |\mu|^2 + \lambda  \xi  \chi  \left(16 |\mu|^4 -16 |\mu|^2 + 3\right)\bigg)\bigg]	\;,\\
	\sigma^\lambda_{\xi\zeta}
	&=& S^\lambda_\xi /c   - \mathcal{E}^{(0)} \frac{(\theta\rho|\mu|)^2}{2} \theta   (\theta \zeta  \xi +\lambda  \chi )|\mu|^2 \;,\\
	\sigma^\lambda_{\chi\zeta}
	&=&  S^\lambda_{\chi} /c   + \mathcal{E}^{(0)}\lambda \frac{(\theta\rho|\mu|)^2}{2} \theta   (\xi - \lambda \theta \zeta  \chi )|\mu|^2 \;,\\
	\sigma^\lambda_{\zeta\zeta} 
	&=&  E^\lambda - \mathcal{E}^{(0)} (\theta\rho|\mu|)^2\left[ 1 
	+ \left(\frac{\theta\rho|\mu|}{2}\right)^2 2 \left(2|\mu|^2 + (4|\mu|^2-3)\left(1-\rho^2|\mu|^2\right)\right) 	\right] \;,
\end{eqnarray}}
where $\mathcal{E}^{(0)} =c^2\varepsilon_0 w_0^2 E_0^2 |v_0|^2$. 

\section{Validity of the Linear Approximation of General Relativity}
\label{sec:validity}

In the linearized version of general relativity, we decompose the metric into the Minkowski metric plus a perturbation, which is assumed to be small, Eq.~(\ref{krokussli}). In this section, we make a rough calculation - just considering orders of magnitude - to verify that the linear approximation is justified, i.e.~that it is possible to neglect terms quadratic in the metric perturbation. 

From the Einstein equations it follows that the second derivative of the metric perturbation is proportional to $\frac{8\pi G}{c^4}$ times the energy-momentum tensor,
\begin{eqnarray}
	\partial^2 h
	&\sim& \frac{8\pi G}{c^4} T\;.
\end{eqnarray}
When considering spatial components (the other components can be considered to be of the same order of magnitude), we integrate to obtain an area $A$ on the right hand side,
\begin{eqnarray}
	h
	&\sim & \frac{8\pi G}{c^4}TA\;.
\end{eqnarray}
Identifying $TAc$ as the Power $P$, we obtain
\begin{eqnarray}
	h
	&\sim & \frac{8\pi G}{c^5} P\;.
\end{eqnarray}
In our calculation, we wrote the metric perturbation in the form (where we write $\epsilon$ for all expressions of order $O(1)$ in $\theta$)
\begin{eqnarray}
	h 
	&\sim & \varepsilon+\theta\varepsilon+\theta^2\varepsilon +\theta^3\epsilon +\theta^4\epsilon\;.
\end{eqnarray}
The linearized theory is valid if one can neglect terms of the order $O(h^2)$, i.e.~if $h^2\ll h$. In our case, this condition translates to $\epsilon\ll \theta^{4}$. From the above equations, we see that ${\epsilon}\sim\frac{8\pi G}{c^5}P$. The condition then becomes
\begin{eqnarray}
	\frac{8\pi G}{c^5}P
	&\ll& \theta^{4}\;.
\end{eqnarray}
For a power of the order of magnitude $P\sim 10^{15}\;{\rm W}$, we thus have to require
$
	\theta
	\gg 10^{-10}
$.
If we consider $\theta$ to be equal to zero, the condition becomes $\epsilon^2\ll \epsilon$, which is also satisfied.

\vfill
\section{Expansion of Christoffel symbols and curvature tensor}
\label{sec:Rexpansion}

With the Eqs.~(\ref{eq:gammadef}) and (\ref{eq:expansionh}), we can derive a direct relation between the terms of the expansions $(\gamma^{\lambda(n)})^{\bar\alpha}_{\bar\beta\bar\gamma}$ and $h_{ \bar\alpha\bar\beta}^{\lambda(n)}$. We obtain for $\bar i, \bar j,\bar k \in \{\xi,\chi\}$\\
\begin{minipage}{0.5\textwidth}
\begin{eqnarray}
	\A (\gamma^{\lambda(n)})^{\tau}_{\tau\zeta} &=& (\gamma^{\lambda(n)})^{\zeta}_{\tau\tau} = -\frac{1}{2w_0^2}\partial_{\theta\zeta} h^{\lambda(n-1)}_{\tau\tau}\;, \\
	\A (\gamma^{\lambda(n)})^{\zeta}_{\zeta\zeta} &=&  \frac{1}{2w_0^2}\partial_{\theta\zeta} h^{\lambda(n-1)}_{\zeta\zeta}\;,\\
	\A (\gamma^{\lambda(n)})^{\tau}_{\zeta\zeta} &=&  -\frac{1}{w_0^2} \partial_{\theta\zeta} h^{\lambda(n-1)}_{\zeta\tau}\\
	\A (\gamma^{\lambda(n)})^{\bar j}_{\zeta\zeta} &=&  -\frac{1}{2w_0^2}\partial_{\bar j} h^{\lambda(n)}_{\zeta\zeta} + \frac{1}{w_0^2}\partial_{\theta\zeta} h^{\lambda(n-1)}_{\bar j\zeta}\;,\\
	\A (\gamma^{\lambda(n)})^{\bar j}_{\zeta\tau} &=&  -\frac{1}{2w_0^2}\partial_{\bar j} h^{\lambda(n)}_{\zeta\tau} + \frac{1}{2w_0^2}\partial_{\theta\zeta} h^{\lambda(n-1)}_{\bar j\tau}\;,\\
	\A (\gamma^{\lambda(n)})^{\bar j}_{\tau\tau} &=&  -\frac{1}{2w_0^2}\partial_{\bar j} h^{\lambda(n)}_{\tau\tau}\;,\\
	\A (\gamma^{\lambda(n)})^{\zeta}_{\bar j\zeta} &=& \frac{1}{2w_0^2}\partial_{\bar j} h^{\lambda(n)}_{\zeta\zeta}\;, 
\end{eqnarray}
\end{minipage}
\begin{minipage}{0.5\textwidth}
\begin{eqnarray}\label{eq:gammaexprelation}
	\A (\gamma^{\lambda(n)})^{\zeta}_{\bar i\bar j} &=&  \frac{1}{2w_0^2}\left(\partial_{\bar i} h^{\lambda(n)}_{\bar j\zeta} + \partial_{\bar j} h^{\lambda(n)}_{\bar i\zeta} - \partial_{\theta\zeta} h^{\lambda(n-1)}_{\bar i\bar j}\right)\;,\\
	\A (\gamma^{\lambda(n)})^{\zeta}_{\bar j\tau} &=&  \frac{1}{2w_0^2}\left(\partial_{\bar j} h^{\lambda(n)}_{\tau\zeta} - \partial_{\theta\zeta} h^{\lambda(n-1)}_{\bar j\tau}\right)\;,\\
	\A (\gamma^{\lambda(n)})^{\tau}_{\bar j\zeta} &=&  -\frac{1}{2w_0^2}\left(\partial_{\bar j} h^{\lambda(n)}_{\tau\zeta} + \partial_{\theta\zeta} h^{\lambda(n-1)}_{\bar j\tau}\right)\;,\\
	\A (\gamma^{\lambda(n)})^{\tau}_{\bar i\bar j} &=&  -\frac{1}{2w_0^2}\left(\partial_{\bar i} h^{\lambda(n)}_{\tau \bar j} + \partial_{\bar j} h^{\lambda(n)}_{\tau \bar i}\right)\;,\\
	\A (\gamma^{\lambda(n)})^{\bar i}_{\bar j\tau} &=&  \frac{1}{2w_0^2}\left(\partial_{\bar j}
	 h^{\lambda(n)}_{\tau\bar i} - \partial_{\bar i} h^{\lambda(n)}_{\tau \bar j}\right)\;,\\
	 \A (\gamma^{\lambda(n)})^{\bar i}_{\zeta\bar j} &=&  \frac{1}{2w_0^2}\left(\partial_{\bar j} h^{\lambda(n)}_{\bar i\zeta} - \partial_{\bar i} h^{\lambda(n)}_{\bar j\zeta} + \partial_{\theta\zeta} h^{\lambda(n-1)}_{\bar i\bar j}\right)\;,\\
	 (\gamma^{\lambda(n)})^{\bar k}_{\bar i\bar j} &=&  \frac{1}{2w_0^2}\left(\partial_{\bar i} h^{\lambda(n)}_{\bar j\bar k} + \partial_{\bar j} h^{\lambda(n)}_{\bar i\bar k} - \partial_{\bar k} h^{\lambda(n)}_{\bar i\bar j}\right)\;,
\end{eqnarray}
\end{minipage}\\

where $h_{\bar\alpha\bar\beta}^{ \lambda(n)}=0$ if $n<0$.
With the Eqs.~(\ref{eq:R}) and (\ref{eq:expansionh}), we can derive a direct relation between the terms of the expansions $r^{\lambda(n)}_{\bar\alpha\bar\beta\bar\gamma\bar\delta}$ and $h_{ \bar\alpha\bar\beta}^{\lambda(n)}$. With $\bar j,\bar k \in \{\xi,\chi\}$, we obtain

\begin{minipage}{0.5\textwidth}
\begin{eqnarray}\label{eq:Rexprelation}
	\A r^{\lambda(n)}_{\xi\chi\xi\chi} &=&  \partial_\xi\partial_\chi h^{\lambda(n)}_{\xi\chi}
		-\frac{1}{2}\left(\partial^2_\xi h^{\lambda(n)}_{\chi\chi} + \partial^2_\chi h^{\lambda(n)}_{\xi\xi}\right)\;,\\
	\A r^{\lambda(n)}_{\bar j\zeta \bar j\bar k} &=&  \frac{1}{2}\partial_{\theta\zeta}\left(\partial_{\bar j} h^{\lambda(n-1)}_{\bar j\bar k} - \partial_{\bar k} h^{\lambda(n-1)}_{\bar j\bar j}\right)
		-\frac{1}{2}\partial_{\bar j}\left(\partial_{\bar j} h^{\lambda(n)}_{\zeta\bar k} - \partial_{\bar k} h^{\lambda(n)}_{\zeta \bar j}\right)\;,\\
		\A r^{\lambda(n)}_{\bar j \zeta \bar k \zeta} &=&  \frac{1}{2}\left(\partial_{\theta\zeta}\partial_{\bar k} h^{\lambda(n-1)}_{\bar j \zeta} - \partial^2_{\theta\zeta} h^{\lambda(n-2)}_{\bar j\bar k} + \partial_{\theta\zeta}\partial_{\bar j} h^{\lambda(n-1)}_{\bar k \zeta}  - \partial_{\bar j}\partial_{\bar k} h^{\lambda(n)}_{\zeta \zeta}\right)\;,\\
		\A r^{\lambda(n)}_{\bar j \tau \bar j\bar k} &=&  \frac{1}{2}\partial_{\bar j}\left(\partial_{\bar k} h^{\lambda(n)}_{\tau \bar j} - \partial_{\bar j} h^{\lambda(n)}_{\tau\bar k}\right)\;,\\
		\A r^{\lambda(n)}_{\bar j \tau \bar k \tau} &=&  -\frac{1}{2}\partial_{\bar j}\partial_{\bar k} h^{\lambda(n)}_{\tau\tau} \;,
		\end{eqnarray}
\end{minipage}
\begin{minipage}{0.5\textwidth}
\begin{eqnarray}
		\A r^{\lambda(n)}_{\bar j \tau \zeta \tau} &=&  -\frac{1}{2}\partial_{\bar j}\partial_{\theta\zeta} h^{\lambda(n-1)}_{\tau\tau}\;, \\
		\A r^{\lambda(n)}_{\bar j \tau \zeta \bar k} &=&  \frac{1}{2}\partial_{\bar j}\left(\partial_{\bar k} h^{\lambda(n)}_{\tau \zeta} - \partial_{\theta\zeta} h^{\lambda(n-1)}_{\tau\bar k}\right)\;,\\
		\A r^{\lambda(n)}_{\zeta \tau \zeta \bar k} &=&  \frac{1}{2}\partial_{\theta\zeta}\left(\partial_{\bar k} h^{\lambda(n-1)}_{ \tau \zeta} - \partial_{\theta\zeta} h^{\lambda(n-2)}_{ \tau\bar k}\right)\;,\\
		 r^{\lambda(n)}_{\zeta \tau \zeta \tau} &=&  -\frac{1}{2}\partial^2_{\theta\zeta} h^{\lambda(n-2)}_{ \tau \tau} \;.
		\end{eqnarray}
\end{minipage}
\vfill

\section{Metric perturbation for large distances between emitter and absorber of the laser beam up to third order in the divergence angle}
\label{sec:mpert}

As stated in Sec.~\ref{sec:metric}, solutions of Eqs.~(\ref{eq:poissonh0}), (\ref{eq:poissonh1}) and (\ref{eq:poissonhhigher}) can be given by Eq.~(\ref{eq:solpoisson}).
However, the Green's function of the Poisson equation in two dimensions is only specified up to a constant which for our degenerate Eq.~(\ref{eq:poissonh0}) in three dimensions becomes a function of $\theta\zeta$. This leads to an additional term $h^{\lambda(n)\rm{rest}}_{\bar\alpha\bar\beta}(\theta\zeta)$ that we have to specify by a further condition. Here, we use the physical condition that the Riemann curvature tensor has to vanish at an infinite distance from the beamline to ensure that no physical effects are induced by the gravitational field of the laser beam at infinity. We find for the general solution
\begin{equation}\label{eq:solpoissongen}
h^{\lambda(n)\rm{gen}}_{\bar\alpha\bar\beta}(\xi,\chi,\theta\zeta)=h^{\lambda(n)}_{\bar\alpha\bar\beta}(\xi,\chi,\theta\zeta) + h^{\lambda(n)\rm{rest}}_{\bar\alpha\bar\beta}(\theta\zeta)\,,
\end{equation}  
where $h^{\lambda(n)}_{\bar\alpha\bar\beta}(\xi,\chi,\theta\zeta)$ is given in Eq.~(\ref{eq:solpoisson}). Since the Riemann curvature tensor is linear in the metric perturbation, it consists of a term induced by $h^{\lambda(n)}_{\bar\alpha\bar\beta}$ and a term induced by $h^{\lambda(n){\rm rest}}_{\bar\alpha\bar\beta}$. The term induced by $h^{\lambda(n)\rm{rest}}_{\bar\alpha\bar\beta}(\theta\zeta)$ does not depend on the distance to the beamline. Let us assume that the term in the curvature tensor induced by the first term in Eq.~(\ref{eq:R}) vanishes for $\rho\rightarrow \infty$. Then, the term in the curvature tensor induced by $h^{\lambda(n)\rm{rest}}_{\bar\alpha\bar\beta}(\theta\zeta)$ has to vanish everywhere for the curvature to vanish for $\rho\rightarrow \infty$. Therefore, $h^{\lambda(n)\rm{rest}}_{\bar\alpha\bar\beta}(\theta\zeta)$ cannot contribute to the curvature tensor and can be set to zero in Eq.~(\ref{eq:solpoissongen}). It turns out that the contribution of the first term in Eq.~(\ref{eq:solpoissongen}) to the curvature tensor vanishes at infinity, indeed, up to the fourth order in $\theta$, as we will show in the following. Therefore, we assume $h^{\lambda(n)\rm{rest}}_{\bar\alpha\bar\beta}(\theta\zeta)=0$ in this article. In the following, we give expressions for $h^{\lambda(n)}_{\bar\alpha\bar\beta}(\xi,\chi,\theta\zeta)$ up to third order in $\theta$.

In zeroth order, we have (see Sec.~\ref{sec:leading} for comparison)
\begin{equation}\label{eq:hzerothorder}
	h^{\lambda(0)}_{\bar\alpha\bar\beta}= \frac{\kappa w_0^2 P_0}{2\pi c}  \left(\frac{1}{2}\rm Ei\left(-2|\mu|^2\rho^2\right)-\log(\rho)\right) \begin{pmatrix}
	1 & 0 & 0 & -1 \\
	0 & 0 & 0 & 0 \\
	0 & 0 & 0 & 0 \\
	-1 & 0 & 0 & 1
	\end{pmatrix}\;.
\end{equation} 
In first order, we have 
\begin{eqnarray}
	h^{\lambda(1)}_{\bar\alpha\bar\beta}
	&=&	\frac{\kappa  P_0 w_0^2 }{8 \pi  c \rho^2}  \left(1-e^{-2 |\mu|^2\rho^2}\right)\begin{pmatrix}
			0& -(\theta \zeta  \xi +\lambda  \chi ) & (\lambda  \xi - \theta \zeta  \chi ) & 0\\
			-(\theta \zeta  \xi +\lambda  \chi ) & 0 & 0 & (\theta \zeta  \xi +\lambda  \chi )\\
			(\lambda  \xi - \theta \zeta  \chi ) & 0 & 0 & -(\lambda  \xi - \theta \zeta  \chi )\\
			0 & (\theta \zeta  \xi +\lambda  \chi ) & -(\lambda  \xi - \theta \zeta  \chi ) & 0
		\end{pmatrix}\;.
\end{eqnarray}
In second order, we find for the only non-vanishing independent components of the metric perturbation
\begin{eqnarray}
	h^{\lambda(2)}_{\tau\tau} &=& \frac{\kappa w_0^2 P_0}{32\pi c}  \left(4\rm Ei\left(-2|\mu|^2\rho^2\right)-8\log(\rho)-(5-(4-6\rho^2)|\mu|^2-8\rho^2|\mu|^4)e^{-2 |\mu|^2\rho^2}\right)\;,\\
	h^{\lambda(2)}_{\tau\zeta} &=& -\frac{\kappa w_0^2 P_0}{32\pi c}  \left(2\rm Ei\left(-2|\mu|^2\rho^2\right)-4\log(\rho)-(3-(4-6\rho^2)|\mu|^2-8\rho^2|\mu|^4)e^{-2 |\mu|^2\rho^2}\right)\;,\\
	h^{\lambda(2)}_{\zeta\zeta} &=& -\frac{\kappa w_0^2 P_0}{32\pi c}  \left(1-(4-6\rho^2)|\mu|^2-8\rho^2|\mu|^4\right)e^{-2 |\mu|^2\rho^2}\;,\\
	\A h^{\lambda(2)}_{\xi\xi} &=& \frac{\kappa w_0^2 P_0}{32\pi\rho^4 |\mu|^2 c}  \Bigg(2\rho^4|\mu|^2 \rm Ei\left(-2|\mu|^2\rho^2\right)- 4\rho^4|\mu|^2\log(\rho) + (\xi^2-\chi^2) - 2(\xi^2-\chi^2 - 2\theta\zeta\lambda\xi\chi)|\mu|^2
	\\
	&&
	+ \left(-(\xi^2-\chi^2) + 2(\xi^2 - \chi^2 -2\theta\zeta\lambda\xi\chi- 2\rho^2\xi^2)|\mu|^2 + 4\rho^2(\xi^2 - \chi^2 -2\theta\zeta\lambda\chi\xi))|\mu|^4\right)  e^{-2 |\mu|^2\rho^2}
	\Bigg)\;,\\
	\A h^{\lambda(2)}_{\chi\chi} &=& \frac{\kappa w_0^2 P_0}{32\pi\rho^4 |\mu|^2 c}  \Bigg(2\rho^4|\mu|^2 \rm Ei\left(-2|\mu|^2\rho^2\right)- 4\rho^4|\mu|^2\log(\rho) - (\xi^2-\chi^2) + 2(\xi^2-\chi^2 - 2\theta\zeta\lambda\xi\chi)|\mu|^2
	\\
	&&
	- \left(-(\xi^2-\chi^2) + 2(\xi^2 - \chi^2 - 2\theta\zeta\lambda\xi\chi + 2\rho^2\chi^2)|\mu|^2 + 4\rho^2(\xi^2 - \chi^2 - 2\theta\zeta\lambda\chi\xi))|\mu|^4\right)  e^{-2 |\mu|^2\rho^2}
	\Bigg)\;,\\
	h^{\lambda(2)}_{\xi\chi} &=& -\frac{\kappa w_0^2 P_0}{16\pi\rho^4 |\mu|^2 c}  \left(1 - (1 + 2\rho^2|\mu|^2)e^{-2 |\mu|^2\rho^2}\right)\left( -\xi\chi + (2\xi\chi +\theta\zeta\lambda(\xi^2-\chi^2))|\mu|^2\right)\,.
\end{eqnarray}
In third order, we obtain the only non-vanishing independent components
\begin{eqnarray}
	\A h^{\lambda(3)}_{\tau\xi} &=& -\frac{\kappa  P_0 w_0^2 }{32 \pi  c \rho^2} \Bigg( (4\theta\zeta\xi + 3\lambda\chi) + \Bigg(-(4\theta\zeta\xi + 3\lambda\chi)-2\rho^2(3\theta\zeta\xi + 2\lambda\chi)|\mu|^2 \\
	 && - 2\rho^2(-2 + 3\rho^2)(\theta\zeta\xi + \lambda\chi)|\mu|^4 + 8\rho^4(\theta\zeta\xi + \lambda\chi) |\mu|^6 \Bigg) e^{-2 |\mu|^2\rho^2}\Bigg)\;,\\
	\A h^{\lambda(3)}_{\tau\chi} &=& -\frac{\kappa  P_0 w_0^2 }{32 \pi  c \rho^2} \Bigg( (4\theta\zeta\chi - 3\lambda\xi) + \Bigg(-(4\theta\zeta\xi - 3\lambda\chi) - 2\rho^2(3\theta\zeta\chi - 2\lambda\xi)|\mu|^2 \\
	&& - 2\rho^2(-2 + 3\rho^2)(\theta\zeta\chi - \lambda\xi)|\mu|^4 + 8\rho^4(\theta\zeta\chi - \lambda\xi) |\mu|^6 \Bigg) e^{-2 |\mu|^2\rho^2}\Bigg) \;,\\
	\A h^{\lambda(3)}_{\zeta\xi} &=& \frac{\kappa  P_0 w_0^2 }{32 \pi  c \rho^2} \Bigg( (2\theta\zeta\xi + \lambda\chi) + \Bigg(-(2\theta\zeta\xi + \lambda\chi)-2\rho^2(2\theta\zeta\xi + \lambda\chi)|\mu|^2 \;\\
	&& - 2\rho^2(-2 + 3\rho^2)(\theta\zeta\xi + \lambda\chi)|\mu|^4 + 8\rho^4(\theta\zeta\xi + \lambda\chi) |\mu|^6 \Bigg) e^{-2 |\mu|^2\rho^2}\Bigg)\;,\\
	\A h^{\lambda(3)}_{\zeta\chi} &=& \frac{\kappa  P_0 w_0^2 }{32 \pi  c \rho^2} \Bigg( (2\theta\zeta\chi - \lambda\xi) + \Bigg(-(2\theta\zeta\chi - \lambda\xi) - 2\rho^2(2\theta\zeta\chi - \lambda\xi)|\mu|^2 \\
	&& - 2\rho^2(-2 + 3\rho^2)(\theta\zeta\chi - \lambda\xi)|\mu|^4 + 8\rho^4(\theta\zeta\chi - \lambda\xi) |\mu|^6 \Bigg) e^{-2 |\mu|^2\rho^2}\Bigg)\,.
\end{eqnarray}
Now, with the expressions for the terms in the expansion of the curvature tensor given in Appendix \ref{sec:Rexpansion}, we can show that the contribution of $h^{\lambda(n)}_{\bar\alpha\bar\beta}(\xi,\chi,\theta\zeta)$ to the curvature vanishes for $\rho\rightarrow\infty$. From Eq.~(\ref{eq:solpoissongen}), we obtain that 
\begin{equation}\label{eq:divdecay}	\partial_{\bar{j}}h^{\lambda(n)}_{\bar\alpha\bar\beta}
	= \frac{2}{4\pi}\int_{-\infty}^\infty d\xi'd\chi' 
\,\frac{x^{\bar j}}{(\xi-\xi')^2+(\chi-\chi')^2}
Q^{\lambda(n)}_{\bar\alpha\bar\beta}(\xi',\chi',\theta\zeta) \,,
\end{equation}
where $x^{\bar{j}}\in \{\xi,\chi\}$. From the expressions in Appendix \ref{sec:pointing}, we see that all terms in the energy-momentum tensor decay like $\exp(-2|\mu|^2\rho^2)$ for $\rho\rightarrow\infty$. From the expressions above, we find that this is true for $\partial^2_{\theta\zeta} h^{\lambda(0)}_{\bar{\alpha}\bar{\beta}}$ and $\partial^2_{\theta\zeta} h^{\lambda(1)}_{\bar{\alpha}\bar{\beta}}$ as well. Furthermore,  $\partial^2_{\theta\zeta} h^{\lambda(2)}_{\bar{\alpha}\bar{\beta}}$ decays at least as $\rho^{-2}$ for $\rho\rightarrow\infty$. Hence, for $n\le 4$ we find that the sources $Q^{\lambda(n)}_{\bar\alpha\bar\beta}$ -- the terms on the right hand side of the differential equations in Eqs.~(\ref{eq:poissonh0}), (\ref{eq:poissonh1}) and (\ref{eq:poissonhhigher}) -- are falling off at least as $\rho^{-2}$ for $\rho\rightarrow \infty$. Therefore, the first derivatives of $h^{\lambda(n)}_{\bar\alpha\bar\beta}(\xi,\chi,\theta\zeta)$ in the directions $\xi$ and $\chi$ will go to zero for $\rho\rightarrow\infty$ for $n\le 4$. From the expressions above, we find that $\partial^2_{\theta\zeta} h^{\lambda(n-2)}_{\tau\tau}$ and $\partial^2_{\theta\zeta} h^{\lambda(n-2)}_{\tau\bar{k}}$ decay like $\exp(-2|\mu|^2\rho^2)$ for $\rho\rightarrow\infty$ for $n\le 4$. Therefore, we find that the contribution of $h^{\lambda(n)}_{\bar\alpha\bar\beta}(\xi,\chi,\theta\zeta)$ to the curvature vanishes for $\rho\rightarrow\infty$ and $n\le 4$. Hence, the term $h^{\lambda(n)\rm{rest}}_{\bar\alpha\bar\beta}(\theta\zeta)$ can be set to zero as argued above.

\section{An exact solution for the infinitely long laser beam with boundary in the paraxial approximation}
\label{sec:exact}

An exact solution for the infinitely long laser  beam in the paraxial approximation, i.e.~for $\theta=0$, is constructed as follows: We make the ansatz of a plane wave metric \cite{bonnor_1969},
\begin{eqnarray}
	ds^2
	&=& w_0^2\left(-d\tau^2+d\xi^2+d\chi^2+d\zeta^2\right)+K\big(d\tau-d\zeta\big)^2\;,
\end{eqnarray}
in the dimensionless coordinates $(\tau,\xi,\chi,\zeta)=(ct,x,y,z)/w_0$. The radius of the beam is supposed to be $a$, such that the energy density $\varrho$ is given by $\varrho w_0^2=T_{\tau\tau}=T_{\zeta\zeta}=-T_{\tau\zeta}$ within this radius, and vanishes outside of it. Then the function $K=K(\xi,\chi)$ in the interior region, for $\rho\leq a$, and in the exterior region, for $a< \rho$, is determined by 
\begin{eqnarray}\label{flips}
	\varrho w_0^2
	&=& -\frac{1}{\kappa w_0^2}\big(\partial_\xi^2+\partial_{\chi}^2\big)K_{\rm int}\;,\\\A
	0
	&=& -\frac{1}{\kappa w_0^2}\big(\partial_\xi^2+\partial_{\chi}^2\big)K_{\rm ext}\;.
\end{eqnarray}
For the laser  beam for $\theta=0$, the energy density is given by $\varrho w_0^2=\mathcal{E}^{(0)}=\frac{2P_0}{\pi c}e^{-2\rho^2}$.
Writing equation (\ref{flips}) in cylindrical coordinates, we obtain
\begin{eqnarray}
	\frac{1}{\rho}\partial_\rho\big( \rho\,\partial_\rho K_{\rm int}\big)
	&=& -\frac{2\kappa P_0 w_0^2}{\pi c}  e^{-2\rho^2}\;,\\\A
	\frac{1}{\rho}\partial_\rho\big(\rho\partial_\rho K_{\rm ext}\big)
	&=& 0\;.
\end{eqnarray}
Integrating twice over $\rho$ leads to
\begin{eqnarray}
	K_{\rm int}(\rho)
	&=& \frac{\kappa P_0 w_0^2}{4\pi c}{\rm Ei}\left(-2\rho^2\right)+C_1\log(\rho)+C_2\;,\\\A
	K_{\rm ext}(\rho)
	&=&	D_1 \log(\rho)+D_2\;,
\end{eqnarray}
where ${\rm Ei(x)}=\gamma+\log(|x|)+i{\rm arg}(x)+x+\frac{x^2}{4}+\frac{x^3}{18}+...$ is the exponential integral. 
For the metric to be finite at $r=0$, we set $C_1=-{\kappa  P_0 w_0^2}/(2\pi c)$, and for the interior solution to match the exterior solution at $r=a$, i.e.~to be continuous and differentiable, we choose $D_2=0$ and $C_2 = \kappa P_0  w_0^2 (2\pi c)^{-1}\left(e^{-2a^2}\log(a)-\frac{1}{2}{\rm Ei}(-2a^2)\right)$, such that the final solution reads
\begin{eqnarray}
	K_{\rm int}
	&=& -\frac{\kappa P_0 w_0^2}{2\pi c}\left(\log(\rho)-\frac{1}{2}{\rm Ei}(-2\rho^2)-e^{-2a^2}\log(a)+\frac{1}{2}{\rm Ei}(-2a^2)\right)\;,\\\A
	K_{\rm ext}
	&=&	-\frac{\kappa P_0 w_0^2}{2\pi c}\left(1-e^{-2a^2}\right)\log(\rho)\;.
\end{eqnarray}
If the beam is infinitely extended in the transverse direction, we are left with an interior solution only which reads
\begin{eqnarray}
	K(\rho)
	&=& -\frac{\kappa P_0 w_0^2}{2\pi c}\left(\log(\rho)-\frac{1}{2}{\rm Ei}\left(-2\rho^2\right)\right)\;.
\end{eqnarray}
The metric may be written as the Minkowski metric plus a small perturbation $h_{\mu\nu}=K(\rho)M_0$, such that the only non-vanishing independent components of the Riemann curvature tensor $R_{\tau i\tau j}= R_{\zeta i\zeta j}=-R_{\tau i\zeta j}
	=	-\frac{1}{2}\partial_i\partial_j K(\rho)$ (for $i,j\in\{\xi, \chi\}$) are given by
\begin{eqnarray}
	R^{\rm int}_{\tau\xi\tau\xi}= R^{\rm int}_{\zeta\xi\zeta\xi}= -R^{\rm int}_{\tau\xi\zeta\xi}
	&=& -\frac{\kappa P_0 w_0^2}{4\pi c}\,\frac{1}{\rho^4}\left((\xi^2- \chi^2)-\left(4\xi^2\rho^2+\xi^2 - \chi^2 \right)e^{-2\rho^2}\right)
	\;,\\\A
	R^{\rm int}_{\tau \chi\tau \chi}= R^{\rm int}_{\zeta \chi \zeta \chi}= -R^{\rm int}_{\tau \chi\zeta \chi}
	&=& \frac{\kappa P_0 w_0^2}{4\pi c}\,\frac{1}{\rho^4}\left((\xi^2 - \chi^2)+\left(4\chi^2\rho^2-\xi^2 + \chi^2 \right)e^{-2\rho^2}\right)
	\;,\\\A
	R^{\rm int}_{\tau\xi\tau\chi}= R^{\rm int}_{\zeta\xi\zeta\chi}= -R^{\rm int}_{\tau\xi\zeta\chi}
	&=& -\frac{\kappa P_0 w_0^2}{2\pi c}\,\frac{\xi\chi}{\rho^4}\left(1-(1+2 \rho^2) e^{-2\rho^2}\right)
	\;,
\end{eqnarray} 
in the interior region and 
\begin{eqnarray}\label{eq:extcurvbonnor}
	R_{\tau\xi\tau\xi}^{\rm ext}= R_{\zeta\xi\zeta\xi}^{\rm ext}= -R^{\rm ext}_{\tau\xi\zeta\xi}
	&=& -\frac{\kappa P_0 w_0^2}{4\pi c}\,\frac{\xi^2 - \chi^2}{\rho^4}\left(1-e^{-2a^2}\right)
	\;,\\\A
	R_{\tau \chi\tau \chi}^{\rm ext}= R_{\zeta \chi \zeta \chi}^{\rm ext}= -R^{\rm ext}_{\tau \chi\zeta \chi}
	&=& \frac{\kappa P_0 w_0^2}{4\pi c}\,\frac{\xi^2 - \chi^2}{\rho^4}\left(1-e^{-2a^2}\right)
	\;,\\\A
	R_{\tau\xi\tau\chi}^{\rm ext}= R_{\zeta\xi\zeta\chi}^{\rm ext}= -R^{\rm ext}_{\tau\xi\zeta\chi}
	&=& -\frac{\kappa P_0 w_0^2}{2\pi c}\,\frac{\xi\chi}{\rho^4}\left(1-e^{-2a^2}\right)\;,
\end{eqnarray}
in the exterior region. We see that the result for the exterior region corresponds to the Riemann curvature tensor for the infinitely thin beam plus a contribution proportional to $e^{-2a^2}$, which vanishes in the limit as $a\rightarrow 0$. The factor 
\begin{equation}
	P_0^a =\frac{1}{2}\pi c\varepsilon_0 E_0^2 w_0^2(1-e^{-2a^2})= c\varepsilon_0 E_0^2 w_0^2 \int_0^{2\pi} d\phi \int_0^a d\rho\, \rho e^{-2\rho^2}  
\end{equation} 
is the total power in the circular region with radius $a$ that contains the source of the gravitation field seen in the exterior region. Therefore, expressing the curvature in the exterior region through the total power $P_0$, we obtain the same result as for the infinitely thin beam. This coincides with the result from Newtonian gravity that the gravitational field outside of a spherical symmetric source distribution does not depend on the radial dependence of its density. 

\vfill
\section{Metric perturbation for small distances between emitter and absorber of the laser beam}
\label{sec:metricpert}

In this appendix we provide the calculations for the metric perturbation for the case of a small distance between the emitter and absorber of the laser beam up to the second order in more detail. In the beginning we calculate the integrals we would need to calculate if the mirrors at $\zeta=\alpha$ and $\zeta=\beta$ were not curved. In a next step we will include the correction for the case when they are curved.
The beam is assumed to be emitted at the location of the wavefront for which $\zeta=\alpha$ on the $\zeta$-axis, propagate along the $\zeta$-axis and be absorbed at the location of the wavefront for which $\zeta=\beta$ on the $\zeta$-axis.  
The mirrors at the emission and absorption are curved such that the phase along them is constant. The phase of the electric field of the Gaussian beam (without the term including the time) is given by
\begin{eqnarray}
	\varphi(\rho,\zeta)
	&=& \theta\zeta(\rho^2-1)+\frac{2}{\theta}\zeta + {\rm sgn}(\zeta)\frac{\pi}{2} 
		\;.
\end{eqnarray}
For the $\zeta$-coordinate of the mirror at the emission at $\zeta=\alpha$, which we call $\bar\zeta_\alpha$, setting $\varphi(0,\alpha)=\varphi\big(\rho,\bar\zeta_\alpha	(\rho)\big)$, and for the $\bar\zeta_\beta$-coordinate of the mirror at the absorption, setting $\varphi(0,\beta)=\varphi\big(\rho,\bar\zeta_\beta	(\rho)\big)$, we obtain
\begin{eqnarray}
	\bar\zeta_\alpha(\rho)
	&=&	\alpha\left(1-\frac{\theta^2}{2}\rho^2\right)\;,\\\A
	\bar\zeta_\beta(\rho)
	&=& \beta\left(1-\frac{\theta^2}{2}\rho^2\right)\;.
\end{eqnarray}

We start by calculating two integrals that will be useful in the following. The first one is
\begin{eqnarray}
	\mathcal{I}_a
	&=& \int_{-\infty}^\infty d\xi'
	\int_{-\infty}^\infty d\chi '
	\int_{\alpha}^{\beta} d\zeta'	
	\;
	\frac{1}{\sqrt{(\xi-\xi')^2+(\chi -\chi ')^2+(\zeta-\zeta')^2}}
	e^{-2(\xi'{}^2+\chi '{}^2)}\;.
\end{eqnarray}
Introducing for any coordinate $u\in\{\xi,\chi,\zeta\}$ a shifted coordinate $u''$ by $u\;'' = u' - u$, and changing to cylindrical coordinates $(\xi'',\chi '',\zeta'') = (\rho''\cos(\phi''), \rho''\sin(\phi''), z'')$, we obtain, with $\alpha''=\alpha-\zeta$ and $\beta''=\beta-\zeta$,
\begin{eqnarray}
	\mathcal{I}_a
	&=&	
		\int_{\alpha''}^{\beta''}d\zeta''\int_0^{2\pi}d\phi''
		\int_0^\infty d\rho''
		\frac{\rho''}{\sqrt{\rho''{}^2+\zeta''{}^2}}
		e^{-2(\rho''{}^2+\rho^2 + 2\rho\rho''\cos(\phi''-\phi))}\;.
\end{eqnarray}		
Using the Bessel function of the first kind, $J_0(ix)=\frac{1}{\pi}\int_0^\pi d\phi e^{x\cos(\phi)}$, leads to
\begin{eqnarray}\label{g0}
	\mathcal{I}_a
	&=&	2\pi e^{-2\rho^2}
		\int_0^\infty d\rho''\; \rho'' \log\left(\frac{\beta-\zeta+\sqrt{(\beta-\zeta)^2+\rho''{}^2}}{\alpha-\zeta+\sqrt{(\alpha-\zeta)^2+\rho''{}^2}}\right)J_0\left(i4\rho\rho''\right)
		e^{-2\rho''{}^2}\;.
\end{eqnarray}
The second integral we calculate is the same as before but with a factor $\zeta'$ in the nominator,
\begin{eqnarray}
	\mathcal{I}_b
	&=& \int_{\alpha}^\beta d\zeta'	\int_{-\infty}^\infty d\xi'
	\int_{-\infty}^\infty d\chi '\;
	\frac{\zeta'}{\sqrt{(\xi-\xi')^2+(\chi -\chi ')^2+(\zeta-\zeta')^2}}
	e^{-2(\xi'{}^2+\chi '{}^2)}\;.
\end{eqnarray}
In the same way as before, we obtain
\begin{eqnarray}
	\mathcal{I}_b
	&=& \zeta \;\mathcal{I}_a
		+2\pi e^{-2\rho^2} \int_0^\infty d\rho''\;\rho''
		\left(\sqrt{\rho''{}^2+(\beta-\zeta)^2}
		-\sqrt{\rho''{}^2+(\alpha-\zeta)^2}\right)
		J_0(i4\rho\rho'')
		e^{-2\rho''{}^2}.
\end{eqnarray}
The third integral we define is
\begin{eqnarray}
	\mathcal{I}_c
	&=&	\int_{-\infty}^\infty d\xi'\int_{-\infty}^\infty d\chi '
		\int_{\alpha}^{\beta}
		d\zeta'\;
		\frac{\zeta'{}^2}{\sqrt{(\xi-\xi')^2+(\chi -\chi ')^2
		+(\zeta-\zeta')^2}}
		e^{-2(\xi'^2+\chi '^2)}\;.
	\end{eqnarray}		
Every other integral we need to solve to calculate the metric perturbation can be expressed through derivatives of these integrals, using the following identities (and equivalently for $\mathcal{I}_b$ or $\mathcal{I}_c$):
{\small
\begin{eqnarray}\label{nebel}
	\int_{\alpha}^\beta d\zeta'	\int_{-\infty}^\infty d\xi'
	\int_{-\infty}^\infty d\chi '\;
	\frac{\xi'}{\sqrt{(\xi-\xi')^2+(\chi -\chi ')^2+(\zeta-\zeta')^2}}
	e^{-2(\xi'{}^2+\chi '{}^2)}\\\A
	= \int_{\alpha''}^{\beta''}d\zeta''\int_{-\infty}^\infty d\xi''\int_{-\infty}^\infty d\chi ''\;
		\frac{\xi''+\xi}{\sqrt{\xi''^2+\chi ''^2+\zeta''^2}}
		e^{-2(\xi''+\xi)^2+(\chi ''+\chi )^2}
	&=&	-\frac{1}{4}\partial_\xi \mathcal{I}_a\;,\\\A
	\int_{\alpha}^\beta d\zeta'	\int_{-\infty}^\infty d\xi'
	\int_{-\infty}^\infty d\chi '\;
	\frac{\xi'^2}{\sqrt{(\xi-\xi')^2+(\chi -\chi ')^2+(\zeta-\zeta')^2}}
	e^{-2(\xi'{}^2+\chi '{}^2)}
	&=& \frac{1}{4}\left(1+\frac{1}{4}\partial_\xi^2\right)\mathcal{I}_a\;,\\\A
	\int_{\alpha}^\beta d\zeta'	\int_{-\infty}^\infty d\xi'
	\int_{-\infty}^\infty d\chi '\;
	\frac{\xi'^4}{\sqrt{(\xi-\xi')^2+(\chi -\chi ')^2+(\zeta-\zeta')^2}}
	e^{-2(\xi'{}^2+\chi '{}^2)}
	&=& \frac{1}{16}\left(\frac{1}{16}\partial_\xi^4+\frac{3}{2}\partial_\xi^2+3\right)\mathcal{I}_a\;.
\end{eqnarray}}
Including the correction of the boundaries of the integral due to the curvature of the mirrors, we obtain for the first integral
\begin{eqnarray}
	\mathcal{I}_A
	&=& \int_{-\infty}^\infty d\xi'
	\int_{-\infty}^\infty d\chi '
	\int_{\bar\zeta_\alpha(\rho')}^{\bar\zeta_\beta(\rho')} d\zeta'	
	\;
	\frac{1}{\sqrt{(\xi-\xi')^2+(\chi -\chi ')^2+(\zeta-\zeta')^2}}
	e^{-2(\xi'{}^2+\chi '{}^2)}\\\A
	&=& -	\int_{-\infty}^\infty d\xi'\int_{-\infty}^\infty d\chi'\;
		\log\left(\frac{\zeta-\bar\zeta_\beta(\rho')+\sqrt{(\zeta-\bar\zeta_\beta(\rho'))^2+(\xi-\xi')^2+(\chi -\chi ')^2}}
		{\zeta-\bar\zeta_\alpha(\rho')+\sqrt{(\zeta-\bar\zeta_\alpha(\rho'))^2+(\xi-\xi')^2+(\chi -\chi ')^2}}\right)
		e^{-2(\xi'^2+\chi '^2)}
	\;.
\end{eqnarray}
Inserting $\bar\zeta_\beta(\rho')=\beta\left(1-\frac{\theta^2}{2}(\xi'^2+\chi '^2)\right)$ and $\bar\zeta_\alpha(\rho')=\alpha\left(1-\frac{\theta^2}{2}(\xi'^2+\chi '^2)\right)$ and expanding to the second order of $\theta$ leads to
\begin{eqnarray}
	\mathcal{I}_A
	&=&	\mathcal{I}_a+\theta^2\delta\mathcal{I}_a\;,
\end{eqnarray}
where we defined
\begin{eqnarray}
	\delta\mathcal{I}_a
	&=&-\frac{1}{2}\int_{-\infty}^\infty d\xi'\int_{-\infty}^\infty d\chi '\;
		(\xi'^2+\chi '^2)
		e^{-2(\xi'^2+\chi '^2)}\\\A
	&&	\left(
		\frac{\beta}{\sqrt{(\beta-\zeta)^2+(\xi'-\xi)^2+(\chi '-\chi )^2}}
		-\frac{\alpha}{\sqrt{(\alpha-\zeta)^2+(\xi'-\xi)^2+(\chi -\chi )^2}}
		\right)	
		\;.
\end{eqnarray}
Changing the coordinates as done previously and using equation (\ref{nebel}), we obtain{\small
\begin{eqnarray}
	\delta \mathcal{I}_a
	&=&	-\frac{1}{8}\left( 2 + \frac{1}{4}(\partial_\xi^2+\partial_\chi ^2)\right)\\\A
	&&	\int_0^{2\pi}d\phi''\int_0^\infty d\rho''\;\rho''\left(
		\frac{\beta}{\sqrt{(\beta-\zeta)^2+\rho''^2}}-\frac{\alpha}{\sqrt{(\alpha-\zeta)^2+\rho''^2}}\right)
		e^{-2\big(\rho''^2+\rho^2 + 2\rho\rho''\cos(\phi-\phi'')\big)}\\\A
	&=&	-\frac{\pi}{4}\left( 2 + \frac{1}{4}(\partial_\xi^2+\partial_\chi ^2)\right)
		\left(e^{-2\rho^2}\int_0^\infty d\rho''\;\rho''
		\left(\frac{\beta}{\sqrt{(\beta-\zeta)^2+\rho''^2}}
		-\frac{\alpha}{\sqrt{(\alpha-\zeta)^2+\rho''^2}}\right)
		J_0(i4\rho\rho'')e^{-2\rho''^2}\right)\;,	
\end{eqnarray}}
where we express again the integration over the angle through the Bessel function of the first kind. Adjusting the boundaries in the second integral, we obtain{
\begin{eqnarray}
	\mathcal{I}_B
	&=&	\int_{-\infty}^\infty d\xi'\int_{-\infty}^\infty d\chi '
		\int_{\bar\zeta_\alpha(\rho')}^{\bar\zeta_{\beta}(\rho')}
		d\zeta'\;
		\frac{\zeta'}{\sqrt{(\xi-\xi')^2+(\chi -\chi ')^2
		+(\zeta-\zeta')^2}}
		e^{-2(\xi'^2+\chi '^2)}\\\A
	&=&	\zeta\mathcal{I}_A
		+\int_{-\infty}^\infty d\xi'\int_{-\infty}^\infty d\chi '\;
		e^{-2(\xi'^2+\chi '^2)}\\\A
	&&	\left(\sqrt{(\xi-\xi')^2+(\chi -\chi ')^2+\left(\bar\zeta
		_\beta(\rho')-\zeta\right)^2}
		-\sqrt{(\xi-\xi')^2+(\chi -\chi ')^2+\left(\bar\zeta_\alpha
		(\rho')-\zeta\right)^2}\right)
		\;.	
\end{eqnarray}}
Since the integral $\mathcal{I}_B$ only appears in the second order of the metric perturbation, it is enough to expand it to the lowest order in $\theta$,
\begin{eqnarray}
	\mathcal{I}_B
	&=&	\mathcal{I}_b+O(\theta)\;.\\\A
\end{eqnarray}
The metric perturbation, which is given by integrating over the retarded energy momentum tensor divided by the distance from the observer to the source point,
\begin{eqnarray}
	h_{\alpha\beta}^\lambda
	&=& \frac{4G}{c^4} \int_{-\infty}^\infty d\xi'
		\int_{-\infty}^\infty d\chi '
		\int_{\bar\zeta_\alpha(\rho')}^{\bar\zeta_\beta(\rho')}
		d\zeta'\;
		\frac{T^\lambda_{\alpha\beta}\left(\tau-\sqrt{(\xi-\xi')^2
		+(\chi -\chi ')^2+(\zeta-\zeta')^2},\xi',\chi ',\zeta'\right)}
		{\sqrt{(\xi-\xi')^2
		+(\chi -\chi ')^2+(\zeta-\zeta')^2}}\;,
\end{eqnarray}
is then found to be, expressed in terms of the integrals calculated above,
{\small
\begin{eqnarray}\label{kerze}
	h_{\alpha\beta}^\lambda
	&=&	\frac{\kappa w_0^2 P_0}{{ 2}\pi^2 c} 
	 \left[
			\mathcal{I}_a {\begin{pmatrix}
1&0&0&-1\\0&0&0&0\\0&0&0&0\\-1&0&0&1
\end{pmatrix}				}
			+ \frac{\lambda \theta}{4}
\begin{pmatrix}
0&\partial_\chi &-\partial_\xi & 0\\
\partial_\chi &0&0&-\partial_\chi\\
 -\partial_\xi &0&0&\partial_\xi\\
0 &-\partial_\chi&\partial_\xi&0
\end{pmatrix}					 
			\mathcal{I}_a
\right.  \\\A
	&&		+\frac{\theta^2}{4}\left(
	\Bigg(
	\frac{1}{2}\left(\partial_\xi^2+\partial_\chi^2\right) \mathcal{I}_c+\left(1-\frac{1}{4}\left(\partial_\xi^2+\partial_\chi^2\right) -\frac{2}{ 16^2}\left(\partial_\xi^2+\partial_\chi^2\right)^2\right) \mathcal{I}_a+4\delta\mathcal{I}_a
	\Bigg)
	\begin{pmatrix}
1&0&0&-1\\0&0&0&0\\0&0&0&0\\-1&0&0&1
\end{pmatrix}\right.\\\A
&&\left.\left.+\left(1+\frac{1}{8}\left(\partial_\xi^2+\partial_\chi^2\right)\right) \mathcal{I}_a
\begin{pmatrix}
					2&0&0&-1\\
					0&0&0&0\\0&0&0&0\\
					-1&0&0&0
				\end{pmatrix}
				+\begin{pmatrix}
			0&\partial_\xi&\partial_\chi&0\\
			\partial_\xi&0&0&-\partial_\xi\\
			\partial_\chi&0&0&-\partial_\chi\\
			0&-\partial_\xi&-\partial_\chi&0
\end{pmatrix}						
			\mathcal{I}_b	
			+\begin{pmatrix}
				0&0&0&0\\
				0&1+\frac{1}{4}\partial_\chi ^2 & -\frac{1}{4}\partial_\xi\partial_\chi  & 0\\
				0&-\frac{1}{4}\partial_\xi\partial_\chi  & 1+\frac{1}{4}\partial_\xi^2&0\\
				0&0&0&0
				\end{pmatrix}\mathcal{I}_a
	 \right)\right]
	\;.
\end{eqnarray}}

\vfill
\section{The infinitely thin beam as the limit of small beam waists in the laser  beam}
\label{sec:infbeam}

For $\theta=0$, the condition $\theta \zeta\ll 1$, which is equivalent to $\theta z \ll w_0$, is satisfied also for small beam waists, more specifically for $w_0\ll l_{\rm var}$, where the length scale $l_{\rm var}$ is defined by $l_{\rm var}=\min\left\{ x,y \right\}$. In this case, only the zeroth order of the solution for the laser  beam is non-zero, and one recovers the solution of the infinitely thin beam: Equation (\ref{eq:brombeere}) written in the coordinates $(x,y,z)$ reads (as can be seen from the expression for $\mathcal{I}_a$ in appendix \ref{sec:metricpert})
\begin{eqnarray}
	I^{(0)}
	&=& \frac{\kappa P_0}{2\pi^2 c} \int_a^b dz'\int_{-\infty}^\infty dx'dy'\;
		\frac{1}{\sqrt{(x-x')^2+(y-y')^2+(z-z')^2}}e^{-2\frac{x'{}^2+y'{}^2}{w_0^2}}\;.
\end{eqnarray}
Applying twice the saddle point approximation in the form 
\begin{eqnarray}\label{sp}
	\lim_{N\rightarrow\infty}\int_{-\infty}^\infty dx\; g(x) e^{-Nf(x)}
	&=&	\lim_{N\rightarrow\infty}e^{-Nf(x_0)}g(x_0)\sqrt{\frac{2\pi}{Nf''(x)}}\;,
\end{eqnarray}	
where $x_0$ is a stationnary point of $f$, we obtain
\begin{eqnarray}
	\lim_{w_0\rightarrow 0}I^{(0)}
	&=& \lim_{w_0\rightarrow 0}\frac{\kappa w_0^2 P_0 }{4\pi c}\log\left(\frac{b-z+\sqrt{(b-z)^2+r^2}}{a-z+\sqrt{(a-z)^2+r^2}}\right)\;.
\end{eqnarray}
For small enough $w_0$, we thus have approximately
\begin{eqnarray}
	I^{(0)}
	&=& \frac{\kappa w_0^2 P_0 }{4\pi c}\log\left(\frac{b-z+\sqrt{(b-z)^2+r^2}}{a-z+\sqrt{(a-z)^2+r^2}}\right)\;,
\end{eqnarray}
and, written again in dimensionless coordinates,
\begin{eqnarray}
	I^{(0)}
	&=& \frac{\kappa w_0^2 P_0 }{4\pi c}\log\left(\frac{\beta-\zeta+\sqrt{(\beta-\zeta)^2+\rho^2}}{\alpha-\zeta+\sqrt{(\alpha-\zeta)^2+\rho^2}}\right)\;.
\end{eqnarray}

\vfill
\bibliographystyle{plain}
\bibliography{biblio}

\end{document}